\documentclass[preprint]{aastex}
\usepackage{graphicx}   
\usepackage{verbatim}
\usepackage{bm}
\newcommand{\Ks}{{\ensuremath K_s}}
   
\newcommand{\WHIRCdispersion}{0.227}
\newcommand{\WHIRCdispersionnoHr}{0.164}
\newcommand{\WHIRCdispersionnoHrmpts}{0.138}

\newcommand{\sampleMH}{\ensuremath{-18.314 \pm 0.024}}
\newcommand{\WHIRCMH}{\ensuremath{-18.375 \pm 0.066}}
\newcommand{\BNMH}{\ensuremath{-18.224 \pm 0.028}}
\newcommand{\BNMHfixdm}{\ensuremath{-18.248 \pm 0.030}}
\newcommand{\WVMH}{\ensuremath{-18.317 \pm 0.059}}
\newcommand{\CSPMH}{\ensuremath{-18.376 \pm 0.040}}
\newcommand{\KMH}{\ensuremath{-18.449 \pm 0.056}}

\usepackage{color}

\begin{document} 
 
\title{SweetSpot: Near-Infrared Observations of Thirteen Type Ia Supernovae from a New NOAO Survey Probing the Nearby Smooth Hubble Flow}

\author{
Anja~Weyant\altaffilmark{1}, W.~Michael~Wood-Vasey\altaffilmark{1}, 
Lori~Allen\altaffilmark{2}, 
Peter~M.~Garnavich\altaffilmark{3}, 
Saurabh~W.~Jha\altaffilmark{4}, 
Richard~Joyce\altaffilmark{2},
Thomas~Matheson\altaffilmark{2}
}
  
\altaffiltext{1}{
Pittsburgh Particle physics, Astrophysics, and Cosmology Center (PITT PACC).  
Physics and Astronomy Department, University of Pittsburgh, 
Pittsburgh PA, 15260
}
\altaffiltext{2}{
National Optical Astronomy Observatory,  
950 North Cherry Avenue, 
Tucson, AZ, 85719, USA
}
\altaffiltext{3}{
Department of Physics,
225 Nieuwland Science Hall,
Notre Dame, IN, 46556, USA
}
\altaffiltext{4}{
Department of Physics and Astronomy, 
Rutgers, the State University of New Jersey, 
136 Frelinghuysen Road, 
Piscataway, NJ 08854, USA
}

\email{anw19@pitt.edu}

\keywords{supernova,cosmology}

\date{\today}

\begin{abstract}
We present 13 Type Ia supernovae (SNe~Ia) observed in the restframe near-infrared (NIR) from $0.02<z<0.09$ with the WIYN High-resolution Infrared Camera (WHIRC) on the WIYN 3.5-m telescope.
With only 1--3 points per light curve and a prior on the time of maximum from the spectrum used to type the object we measure an $H$-band dispersion of spectroscopically normal SNe~Ia of \WHIRCdispersionnoHr~mag. 
These observations continue to demonstrate the improved standard brightness of SNe~Ia in $H$-band even with limited data.
Our sample includes two SNe~Ia at $z\sim0.09$, which represent the most distant restframe NIR $H$-band observations published to date. 

This modest sample of 13 NIR SNe~Ia represent the pilot sample for ``SweetSpot'' -- a three-year NOAO Survey program that will observe 144 SNe~Ia in the smooth Hubble flow.
By the end of the survey we will have measured the relative distance to a redshift of $z\sim0.05$ to 1\%.  
Nearby Type Ia supernova (SN~Ia) observations such as these will test the standard nature of SNe~Ia in the restframe NIR, allow insight into the nature of dust, and provide a critical anchor for future cosmological SN~Ia surveys at higher redshift.
\end{abstract}


\section{Introduction}

The discovery of the accelerating expansion of the Universe with Type Ia supernovae (SNe~Ia) \citep{Riess98,Perlmutter99} has sparked a decade-and-a-half of intensive Type Ia supernova (SN~Ia) studies to pursue the nature of dark energy.
High-redshift SN~Ia surveys attempt to measure the equation-of-state parameter to sufficiently distinguish among dark energy models.  
The majority of this work has been focused on standardizing the restframe optical luminosities of SNe~Ia.
The goal of low-redshift surveys has been to both provide the distance anchor for high-redshift relative distance measurements and to better calibrate SNe~Ia as standard candles through improved understanding of SNe~Ia themselves.  

As the amount of available SN~Ia data has grown dramatically, 
systematic uncertainties have come to dominate cosmological distance measurements with SNe~Ia
 \citep{Albrecht06,Astier06,Wood-Vasey07,Kessler09,Sullivan10,Conley11}. 
A well-established systematic affecting SNe~Ia is dust reddening and extinction~\citep[see, for example,][]{Jha07,Conley07,WangX06,Goobar08,Hicken09b,Wang09,Folatelli10,Foley11,Chotard11,Scolnic13}.  
It is difficult to separate the effects of reddening as a result of dust from intrinsic variation in the colors of SNe~Ia.
Unfortunately, most observations of SNe~Ia are made in the rest-frame optical and UV where reddening corrections are large.  

SNe~Ia are superior distance indicators in the near-infrared (NIR),\footnote{In this paper we use the term ``near-infrared'' to refer to observed wavelengths from $1<\lambda<2.5$~$\mu$m.}  
 with more standard peak $JH\Ks$ magnitudes and relative insensitivity to reddening \citep{Meikle00,Krisciunas04a,Krisciunas07} than in the restframe optical passbands traditionally used in SN~Ia distance measurements.
 Additionally, \citet{Krisciunas04a} found that objects that are peculiar at optical wavelengths such as SN~1999aa, SN~1999ac, and SN~1999aw appear normal at infrared wavelengths.  
Although it appears that the 2006bt-like subclass of SNe have normal decline rates and $V$-band peak magnitudes but display intrinsically-red colors and have broad, slow-declining light curves in the NIR similar to super-Chandra SNe~Ia \citep{Foley10, Phillips12}. 

These early results have motivated several efforts to pursue large samples of SNe~Ia observed in the restframe NIR with 1.3--2.5-m telescopes: 
CSP-I,II~\citep{Contreras10,Folatelli10,Stritzinger11,Kattner12}; 
CfA~\citep{Wood-Vasey08}; 
RAISIN~\citep{Kirshner12}.
The results from these projects to date indicate that SNe~Ia appear to be {\em standard} NIR candles 
to $\lesssim0.15$~mag \citep{Wood-Vasey08,Folatelli10,Kattner12}, particularly in the $H$ band. 
NIR observations of SNe~Ia are a current significant focus of nearby studies of SNe~Ia.  
Recent work by \citet{Barone-Nugent12} used 8-m class telescopes 
to observe 12 SNe~Ia in the NIR from $0.03<z<0.08$ and found promising evidence that the $H$-band peak magnitude of SNe~Ia may have a scatter as small $\sigma_H=0.085$~mag.  This work demonstrated the benefit of using larger-aperture telescopes in overcoming the significantly increased background of the night sky in the NIR.
 
In this paper we introduce a new effort to observe SNe~Ia in the NIR in the nearby smooth Hubble flow.  
``SweetSpot'' is a 72-night, three-year NOAO Survey program (2012B-0500) to observe SNe~Ia in $JH\Ks$ using the WIYN 3.5-m telescope and the WIYN High-resolution Infrared Camera (WHIRC). 
Our goal is to extend the rest-frame $H$-band NIR Hubble diagram to $z \sim 0.08$ to 
(1) verify recent evidence that SN~Ia are excellent standard candles in the NIR, particularly in the $H$ band;
(2) test if the recent correlation between optical luminosity and host galaxy mass holds in the NIR; 
(3) improve our understanding of intrinsic colors of SNe~Ia;
(4) study the nature of dust in galaxies beyond our Milky Way;
(5) and provide a standard well-calibrated NIR restframe reference for future higher-redshift supernova surveys.

In this paper we present results from our 2011B pilot proposal.  
In Section~\ref{sec:sample} we discuss our data reduction and present light curves of 13 SNe~Ia. 
To this sample we add data from the literature (Section~\ref{sec:litdata}) and fit the light curves using SNooPy \citep{Burns11}.  
Details of how we perform the fitting are discussed in Section~\ref{sec:analysis}.  
We present our results, including an $H$-band Hubble diagram, in Section~\ref{sec:results}.
We discuss our overall SweetSpot program strategy and goals 
along with future prospects for restframe $H$-band SN~Ia observations 
in Section~\ref{sec:survey},
and conclude in Section~\ref{sec:conclusion}.


\section{The Observation and Processing of the SN~Ia Sample}
\label{sec:sample} 

\subsection{Observations and Sample Selection}
\label{sec:observations}

We were awarded seven nights of National Optical Astronomy Observatory (NOAO) time in 2011B to image SNe~Ia in the NIR using the WIYN 3.5m Observatory at Kitt Peak National Observatory (KPNO) with the WHIRC detector.  
WHIRC \citep{Meixner10} is a NIR imager (0.9--2.5~$\mu$m) with a 3.\arcmin3 field of view and 0.\arcsec1 pixel scale.
The combination of WIYN+WHIRC allows us to observe SNe~Ia out to a redshift of $\sim0.09$.

Three-and-a-half nights of this time were usable; the rest was lost to bad weather.
The light curves presented here thus typically have only 1--3 points in each filter and are sparser than our eventual program goals of 3--10 points per light curve.
Our sample (see Table~\ref{table:datacitation})  was selected from SNe~Ia reported in the IAU Central Bureau for Astronomical Telegrams (CBET)\footnote{\url{http://www.cbat.eps.harvard.edu/cbet/RecentCBETs.html}} and The Astronomers Telegram (ATel)\footnote{\url{http://www.astronomerstelegram.org/}} that were spectroscopically confirmed as Type Ia and were in our preferred redshift range of $0.02 < z < 0.08$.

Our goal is to have the first observation in the light curve within two weeks of maximum.
We are focused on the time from 10--20 days after B-band maximum light as the most standard brightness for SNeIa in the H-band.
Our awarded time is typically scheduled around the full moon and therefore spaced 2-3 weeks apart.  Additionally, there is a lack of targets at the beginning of the season until
searches are up and running.  When we combine weather with these factors,
we find that about 30\% of our light curves from 2011B have their first observation more than 14 days after maximum.

During the first two semesters of our SweetSpot survey, we have been awarded more nights
per semester, more nights occuring later in the semester, and had better weather.
Preliminary results show that we are doing significantly better in obtaining earlier light-curve points, with only 10\% of our light curves having their first observation more than 14 days after B-band maximum light.

Here we present $J$- and $H$-band light curves of the 13 of the 18 SNe~Ia that were sufficiently isolated from the background light of their host galaxy.
We obtained template images for the other 5 supernovae starting in 2012B during our main NOAO Survey program.  The full host-galaxy-subtracted sample will be presented in future work.
A summary of the SNe~Ia presented in this work can be found in Tables \ref{table:datasummaryI} and \ref{table:datasummaryII}.
We describe our data processing in Section~\ref{sec:detrending} and photometric analysis and calibration in Section~\ref{sec:photometry}.

A typical WIYN observation consisted of a 3x3 grid dither pattern with 30\arcsec\ spacing with a 60~s exposure time at each pointing. 
For objects or conditions requiring more total exposure time, we typically executed the dither pattern multiple times with a 5\arcsec\ offset between dither sets.  
Our observations were conducted in both $J$ and $H$ with priority given to $H$.
We obtained calibration images consisting of a set of 10 dome flats with the flat lamp off and another set with the flat lamp on.  We used the WHIRC ``high'' lamps, which are the standard KPNO MR16 halogen lamps with the reflective surface coated with aluminum by the NOAO coatings lab.
We also obtained dark images for monitoring of the dark behavior of the detector, but 
we do not use these dark images in our analysis.

\subsection{Image Processing and Coaddition}
\label{sec:detrending}

\begin{figure}
\plotone{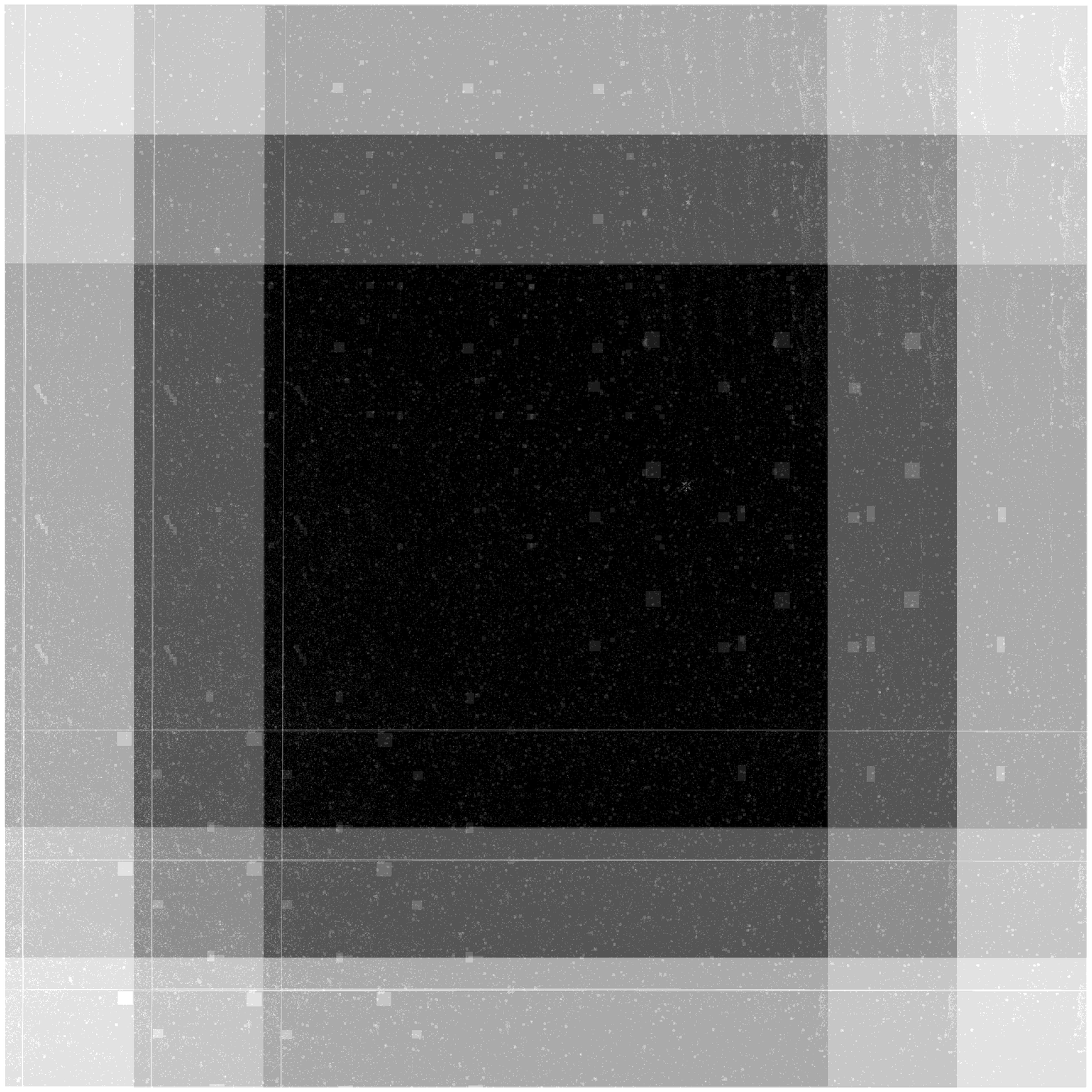}
\caption{Exposure map of a typical stacked observation sequence consisting of a 3x3 grid dither pattern with 30\arcsec\ spacing with a 60~s exposure time at each pointing.}
\label{fig:weightmap}
\end{figure}

\begin{figure}
\plotone{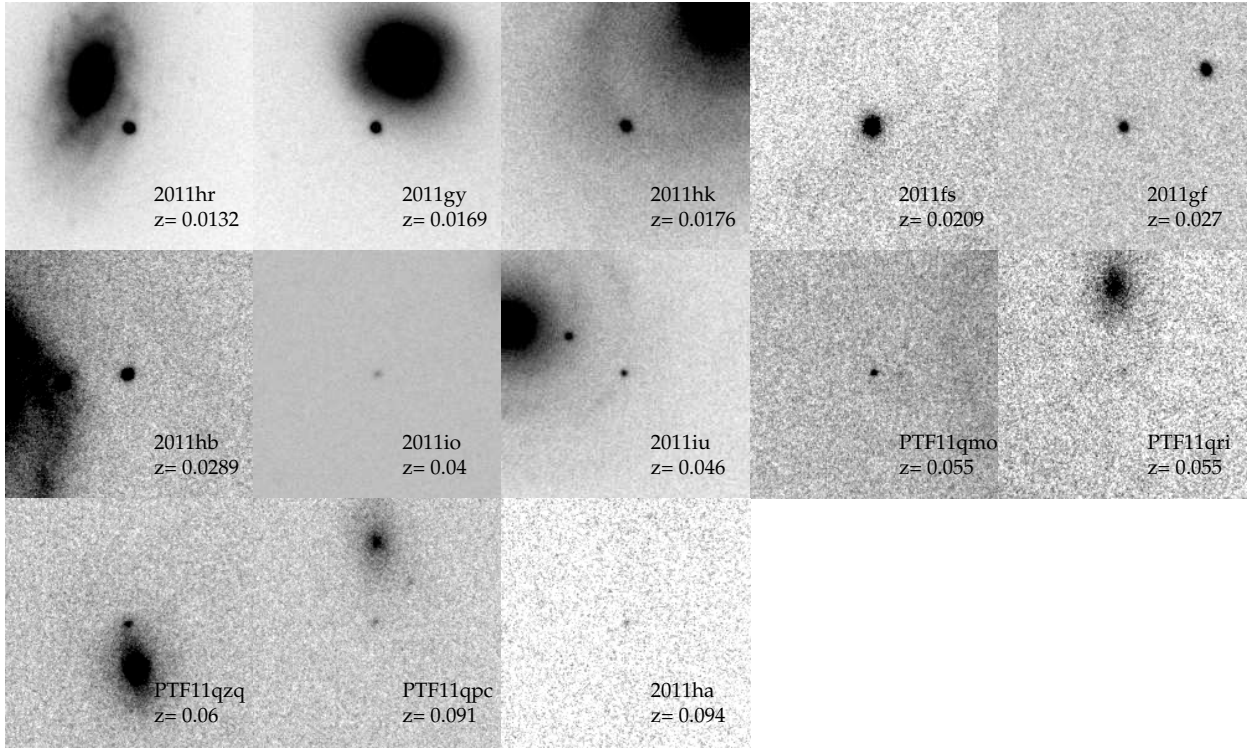}
\caption{Postage stamps of each of the new SNe~Ia presented in this work from our WIYN+WHIRC $H$-band stacked images.  The postage stamps are in order of increasing redshift.  Each image is 10\arcsec square.
}
\label{fig:postagestamps}
\end{figure}

The data were reduced in IRAF\footnote{IRAF is distributed by the National Optical Astronomy Observatory,
which is operated by the Association of Universities for Research
in Astronomy, Inc., under cooperative agreement with the National
Science Foundation} following the steps outlined in the WHIRC Reduction Manual \citep{Joyce09}:
\begin{enumerate}
\item The raw images were trimmed of detector reference pixels outside the main imaging area and corrected for the sub-linear response of the array.
\item The ON dome flats were combined; the OFF dome flats were combined; and the OFF combined dome flat was then subtracted from the ON combined dome flat to yield the pixel-by-pixel response.
\item The pupil ghost (an additive artifact resulting from internal reflection within the optical elements of WHIRC) was removed from this response using the IRAF routine \verb|mscred.rmpupil|.
\item For each target, the set of dithered science images were used to generate a median-filtered sky frame.  The individual science images were then sky-subtracted and flat-fielded using these median frames.  
\item The geometric distortion resulting from a difference in plate scales in the x and y coordinates and field distortion at the input to WHIRC was corrected using the IRAF routine \verb|geotran| and the pre-computed WHIRC geometric distortion calibration from 2009 Mar 05\footnote{\url{http://www.noao.edu/kpno/manuals/whirc/datared.html}}.
\item The individual science images were stacked using the IRAF routine \verb|upsqiid.xyget| to find the common stars in the images and create a registration database between the individual images in an observation sequence.  Intensity offsets were determined from the overlap regions in the registration database and the set of individual images were combined into a composite image using the IRAF routine \verb|upsqiid.nircombine|.
An exposure map of a typical stacked observation sequence can be found in Fig. \ref{fig:weightmap}.
\end{enumerate}
Representative postage stamp images from the processed $H$-band composite images of our supernovae are shown in Fig. \ref{fig:postagestamps}.

\subsection{Photometry and Calibration}
\label{sec:photometry}

We measured the detected counts of the SNe~Ia and the stars in the field with aperture photometry on the stacked images using the Goddard Space Flight Center IDL Astronomy User's Library routines \verb'gcntrd' and \verb'aper'\footnote{\url{http://idlastro.gsfc.nasa.gov/}}.  
We used an aperture diameter of 1.5 FWHM (FWHM values were typically around 2\arcsec) and measured the background in a surrounding sky annulus from 1.5 FWHM + 0.\arcsec1 to 1.5 FWHM + 0.\arcsec6.
These counts in ADU/(60-second) equivalent exposure were converted to instrumental magnitudes $m_{{\rm inst},f}=-2.5\log_{10}{{\rm ADU}/60~{\rm sec}}$.

To calibrate the instrumental magnitudes we first define a transformation between the WHIRC and the Two Micron All Sky Survey \citep[2MASS;][]{2MASS} systems using the following equation
\begin{equation}
 m^{\rm 2MASS}_f -  m_{{\rm inst,}f}^{\rm WHIRC} = {\rm zpt}_f + k_f \left(X-1\right) + c_f \left((m_J^{\rm 2MASS} - m_H^{\rm 2MASS}) - 0.5~{\rm mag}\right) 
\label{eq:transform}
\end{equation}
where $f$ designates the filter, $X$ is the airmass, and the 2MASS color is compared to a reference of $m_J^{\rm 2MASS}-m_H^{\rm 2MASS}=0.5$~mag, which represents the typical color of stars in our fields as well as SNe~Ia after maximum.
We then jointly solve for the zeropoint (${\rm zpt}$), airmass coefficient ($k$)\footnote{Our sign convention for $k$ means that $k$ should be negative.  The opposite convention is also common in the literature.}, and color coefficient ($c$) 
using all instrumental magnitudes measured from 2MASS stars in the fields from our 2011 Nov. 15 and 2012 Jan. 8 nights.  
This procedure was performed separately for each filter following Eq.~\ref{eq:transform}.

\begin{figure}
\epsscale{0.48}
\plotone{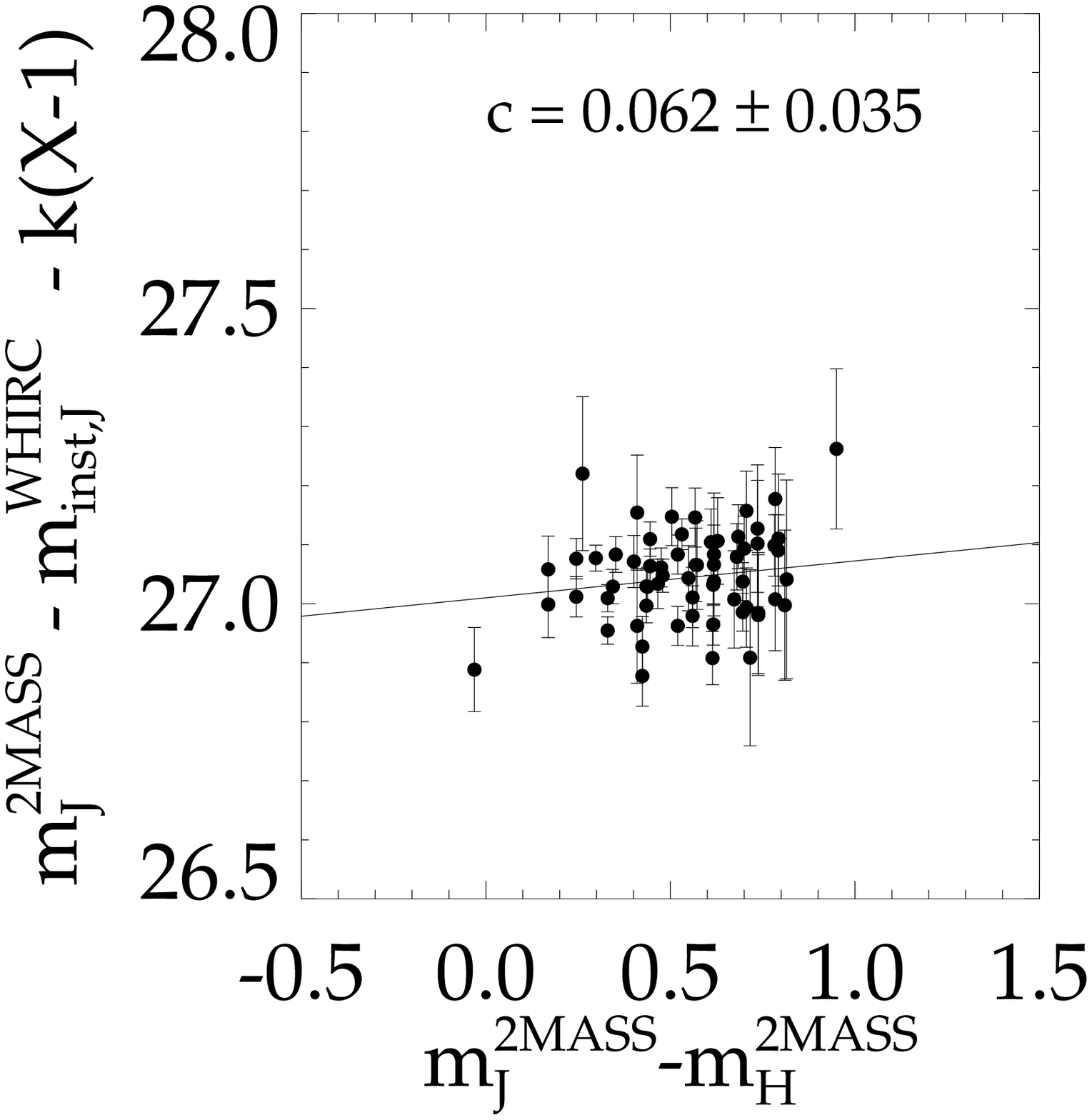}
\plotone{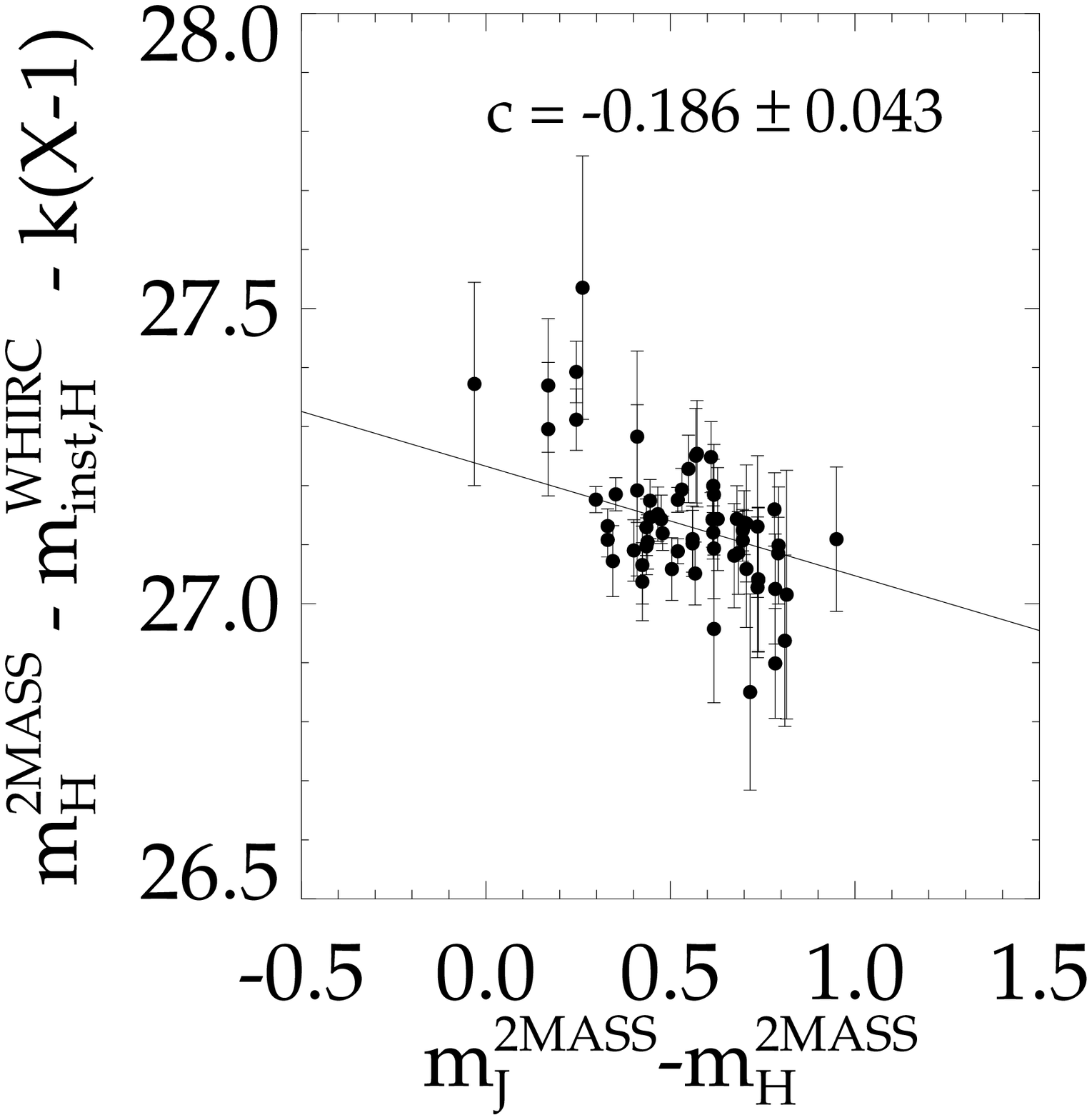}
\caption{The difference in 2MASS magnitude and WHIRC instrumental magnitude corrected for airmass as a function of 2MASS color for the $J$ and $H$ filters.  Fitting Eq. \ref{eq:transform} to these stars (over-plotted) reveals a significant color term between WHIRC and 2MASS. The results of this fit allow us to transform between the WHIRC and 2MASS system and are used to define our natural WHIRC system.}
\label{fig:transform}
\end{figure}

Our fit for each filter is plotted in Fig. \ref{fig:transform} and our fit results are summarized in Table \ref{table:transform}.
We find non-zero color terms of $c_J=0.062\pm0.035$ and $c_H=-0.186\pm0.043$
between the 2MASS and WHIRC systems, and airmass coefficients of $k_J=-0.051\pm0.020$~mag/airmass and $k_H=-0.066\pm0.030$~mag/airmass.
 
\citet{Matheson12} used the same WIYN+WHIRC system to observe the very nearby SN~2011fe in M101, and used ``canonical'' values of $(k_J, k_H, k_\Ks)=(-0.08,-0.04,-0.07)$~mag/airmass (in our sign convention for $k$).  
These values were based on a long-term study of $k_J$, $k_H$, and $k_K$ at KPNO in the 1980s using single-channel NIR detectors.  This effort found a range of values of $-0.12<k_J<-0.07$~mag/airmass, $-0.08<k_H<-0.04$~mag/airmass, and $-0.11<k_K<-0.07$~mag/airmass with a significant seasonal variation dependent on the precipitable water vapor (R. R. Joyce and R. Probst, private communication).
The filters used in these measurements were wider than the standard 2MASS filters or WHIRC filters we use here.
The narrow WHIRC filters do not include some of the significant water-vapor absorption regions included in the NIR filters used in the 1980s KPNO study and thus would reasonably be expected to have a smaller absolute value of $k_J$.
Our determined $k_J$ and $k_H$ values are thus consistent with these previous results.  However, the variation of $k$ in the NIR as a result of water vapor strongly motivates future improvements in tracking precipitable water vapor and NIR extinction to improve the instantaneous determination of $k$.

We then selected a star in each field that was near the supernova and had a similar color to the supernova at the time of our observations. 
These reference stars are listed in Table \ref{table:calibrationstars}.  We used the best observation of the reference star, our fit results from Table \ref{table:transform}, and Eq. \ref {eq:transform} to create a list of calibrated standard stars in the WHIRC natural system.   
We note that our only observation of SN~2011io was taken under partial clouds.
For a given field, the standard star was then used to find the zeropoint for each stacked image as follows  
\begin{equation}
{\rm zpt}_{f,i} = m_{{\rm cal},f}^{\rm WHIRC} - m_{{\rm inst},f,i}^{\rm WHIRC}
\end{equation}
where the $i$ subscript indicates stacked image and $m_{\rm cal}$ is the calibrated standard star for that field. 
This zeropoint was then applied to the measured instrument magnitude from the supernovae to generate the calibrated supernova magnitude in the WHIRC natural system.  
These light curves are presented in Table \ref{table:lightcurves}.

We report magnitudes in the WIYN+WHIRC natural system.\footnote{For reference, the filter transmissions for WIYN+WHIRC can be found at \url{http://www.noao.edu/kpno/manuals/whirc/filters.html}}  

\section{SN~Ia Sample from the Literature}
\label{sec:litdata}
 
To our sample of WHIRC SNe~Ia we add the following data from the literature:  
\begin{itemize}
\item A compilation of 23 SNe~Ia from \citet{Jha99}, \citet{Hernandez00}, \citet{Krisciunas00}, \citet{Krisciunas04a}, \citet{Krisciunas04b}, \citet{Phillips06}, \citet{Pastorello07a}, \citet{Pastorello07b}, and \citet{Stanishev07}.  
This is the same set that was used as the ``literature'' sample by \citet{Wood-Vasey08}.  We use 22 SNe~Ia from this set, one of which was observed by the Carnegie Supernova Project (CSP).  We refer to the 21 SNe~Ia that are unique to this sample as K+ in recognition of the substantial contributions by Kevin Krisciunas to this sample and the field of NIR SNe~Ia.
\item \citet{Wood-Vasey08} presented $JH\Ks$
measurements of 21 SNe~Ia from the Center for Astrophysics (CfA) Supernova Program using the robotic 1.3~m Peters Automated Infrared Imaging Telescope (PAIRITEL; \citealt{Bloom06}) at Mount Hopkins, Arizona.  We use 17 SNe~Ia from this sample which we refer to as WV08.
\item \cite{Contreras10} and \citet{Stritzinger11} present 69 SNe~Ia from the CSP using observations at the Las Campanas Observatory in Chile \citep{Hamuy06}. 
The CSP observations in $YJH\Ks$ were carried out with the Wide Field Infrared Camera (WIRC) attached to the du Pont 2.5~m Telescope and RetroCam on the Swope 1-m telescope supplemented by occasional imaging with the PANIC NIR imager \citep{Osip04} on the Magellan Baade 6.5-m telescope.    We use 55 SNe~Ia from this sample, 6 of which are also in WV08.  We refer to the 49 SNe~Ia that were not observed by \citet{Wood-Vasey08} as CSP. 
\item \citet{Barone-Nugent12} extended the rest-frame NIR sample out to $z\sim0.08$ with 12 SNe~Ia observed in $JH$ on Gemini Observatory's 8.2m Gemini North with the NIR Imager and Spectrometer \citep{Hodapp00} 
and on ESO's 8.1m VLT using HAWK-I \citep{Casali06}.  We use these 12 SNe~Ia and refer to this set as BN12.
\end{itemize}

\begin{figure}
\epsscale{1.0}
\plotone{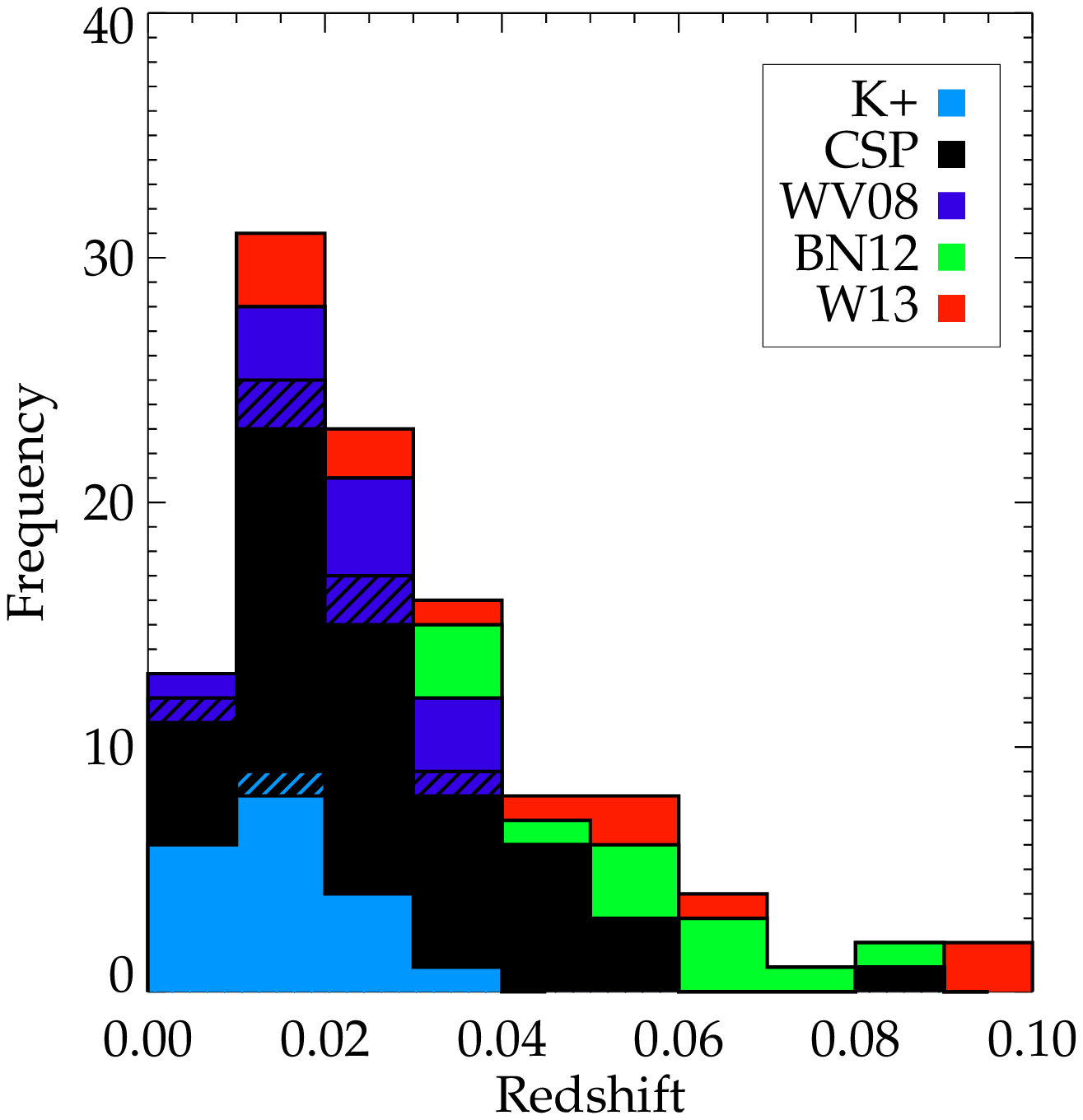}
\caption{Cumulative distribution in redshift of supernovae from the K+ sample in cyan, \citet{Contreras10} and \citet{Stritzinger11} in black (CSP), \citet{Wood-Vasey08} in blue (WV08), \citet{Barone-Nugent12} in green (BN12), and this present paper in red (W13). 
The hatched region represents SN observed by multiple groups.  With WIYN+WHIRC we can probe a large redshift range and populate the NIR Hubble diagram above $z>0.03$ where measurements of the distance-redshift relation are less affected by peculiar velocities.
}
\label{fig:redshiftdistribution}
\end{figure}

To arrive at these samples we removed supernovae that were reported to have a spectrum similar to the sub-luminous SN~1991bg (SN~2006bd, SN~2007N, SN~2007ax, SN~2007ba, SN~2009F); 
were reported to have a spectrum that was peculiar (SN~2006bt, SN~2006ot); 
were identified as possible super-Chandrasekhar mass objects (SN~2007if, SN~2009dc); 
were determined to be highly reddened (SN~1999cl, SN~2003cg, SN~2005A, SN~2006X); 
or were found to have a decline rate parameter $\Delta m_{15} > 1.7$ (SN~2005bl, SN~2005ke, SN~2005ku, SN~2006mr) according to the information provided in \citet{Folatelli10, Contreras10, Stritzinger11, Burns11}. 
We also removed SN~2002cv that \citet{Elias-Rosa08} find to be heavily obscured and SN~2007hx whose photometry is unreliable (Maximilian Stritzinger, private communication).
A redshift histogram of this entire sample, which represents the currently available collection of published normal NIR SNe~Ia, is plotted in Fig. \ref{fig:redshiftdistribution}. 
Note that with WIYN+WHIRC we can reach out to $z\sim0.09$ and cover the entirety of the nearby smooth Hubble flow from $0.03<z<0.08$.

We used the quoted system transmission function reported by each survey.
For SNe~Ia that were observed by multiple surveys, we fit all of the available photometry for the SN~Ia.  

 
\section{Analysis}\label{sec:analysis}

We fit the light-curves using the suite of supernova analysis tools developed by CSP called SNooPy \citep{Burns11}.
We fit the data using SNooPy (version 2.0-267) ``max\_model'' fitting that uses the following model $m_X$:
\begin{eqnarray}
m_X(t-t_{\rm max}) &=& T_Y((t^{\prime}-t_{\rm max})/(1+z),\Delta m_{15}) + m_Y + R_XE(B-V)_{\rm Gal} + \nonumber \\
&& K_{X,Y}(z,(t^{\prime}-t_{\rm max})/(1+z),E(B-V)_{\rm host}, E(B-V)_{\rm Gal})
\end{eqnarray}
where $t$ is time in days in the observer frame, $T_Y$ is the SNooPy light-curve template, $m_Y$ is the peak magnitude in filter Y, $t_{\rm max}$ is the time of maximum in the $B$ band, $\Delta m_{15}$ is the decline rate parameter \citep{Phillips93}, $E(B-V)_{\rm Gal}$ and $E(B-V)_{\rm host}$ are the reddening resulting from the Galactic foreground and the host galaxy, $R_X$ is the total-to-selective absorption for filters $X$, and $K_{X,Y}$ is the cross-band K-correction from rest-frame X to observed $Y$.
The free parameters in this model are $t_{\rm max}$, $\Delta m_{15}$, and $m_Y$.
We do not assume any relationship between the different filters and therefore do not apply any color correction.  We generate the template $T(t, \Delta m_{15})$ from the code of \citet{Burns11} which generates rest-frame templates for $J$ and $H$ from the CSP data \citep{Folatelli10}.  

\begin{figure}
\epsscale{1.0}
\plotone{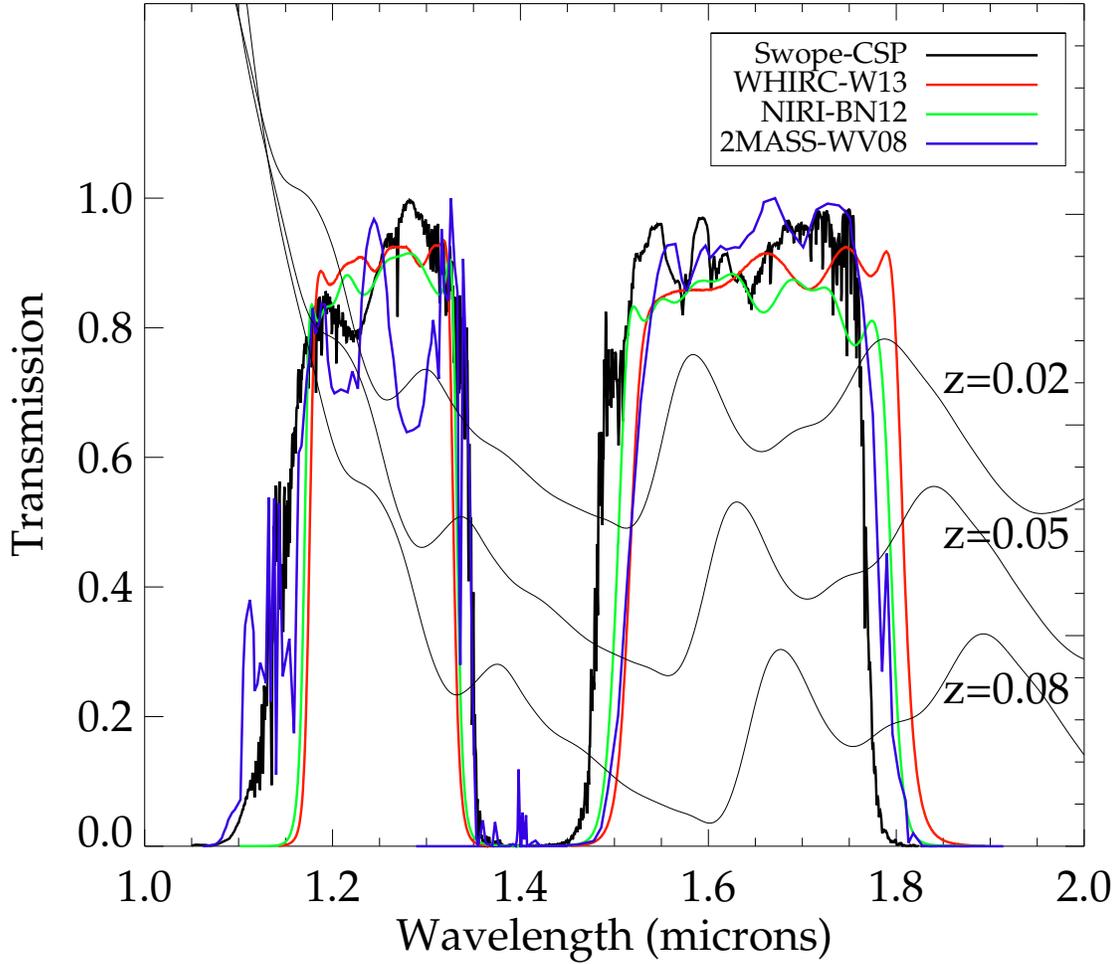}
\caption{Filter transmission for the different instruments in our sample. The atmosphere is included in the filter transmission curve for 2MASS and Swope, but not in the ones for WHIRC and NIRI.  Over-plotted is a synthetic spectrum for a Type Ia which is 30 days old from \citet{Hsiao07} at three different redshifts.  
Note in particular the variation in the red edge of the filters for the different telescope+detector systems and the shifting of a significant NIR feature (restframe $\lambda\sim1.75$~$\mu$m) from $z=0.02$ to $z=0.08$.
}
\label{fig:filterresponseH}
\end{figure}

We use SNooPy to perform the K-corrections on all of the data using the \citet{Hsiao07} spectral templates.  We do not warp or ``mangle'' the spectral template to match the observed color when performing the K-corrections.  A simpler approach makes sense as we are interested in measuring the peak brightness using one NIR band and a prior on $t_{\rm max}$.  
In Fig.~\ref{fig:filterresponseH} we plot the $H$-band filter transmission for the different surveys in our sample.  Overlaid are synthetic spectra at various redshifts.  Note the difference in widths and up to 0.05~$\mu$m shift in the positions of the blue and red edges of the different $H$-band filters.  While SNe~Ia are standard in their rest-frame $H$-band brightness, there is a significant feature at 1.8~$\mu$m which moves longward of the red edge of the $H$-band filter quickly from just $z=0$ to $z=0.05$.  This feature means that it is quite important to have well-understood transmission functions and spectral templates.  However, given that the main effect is the feature moving across the edge of the filter cutoff, knowing the filter bandpass provides most of the necessary information without an immediate need for a full system transmission function.

\begin{figure}
\epsscale{0.25}
\plotone{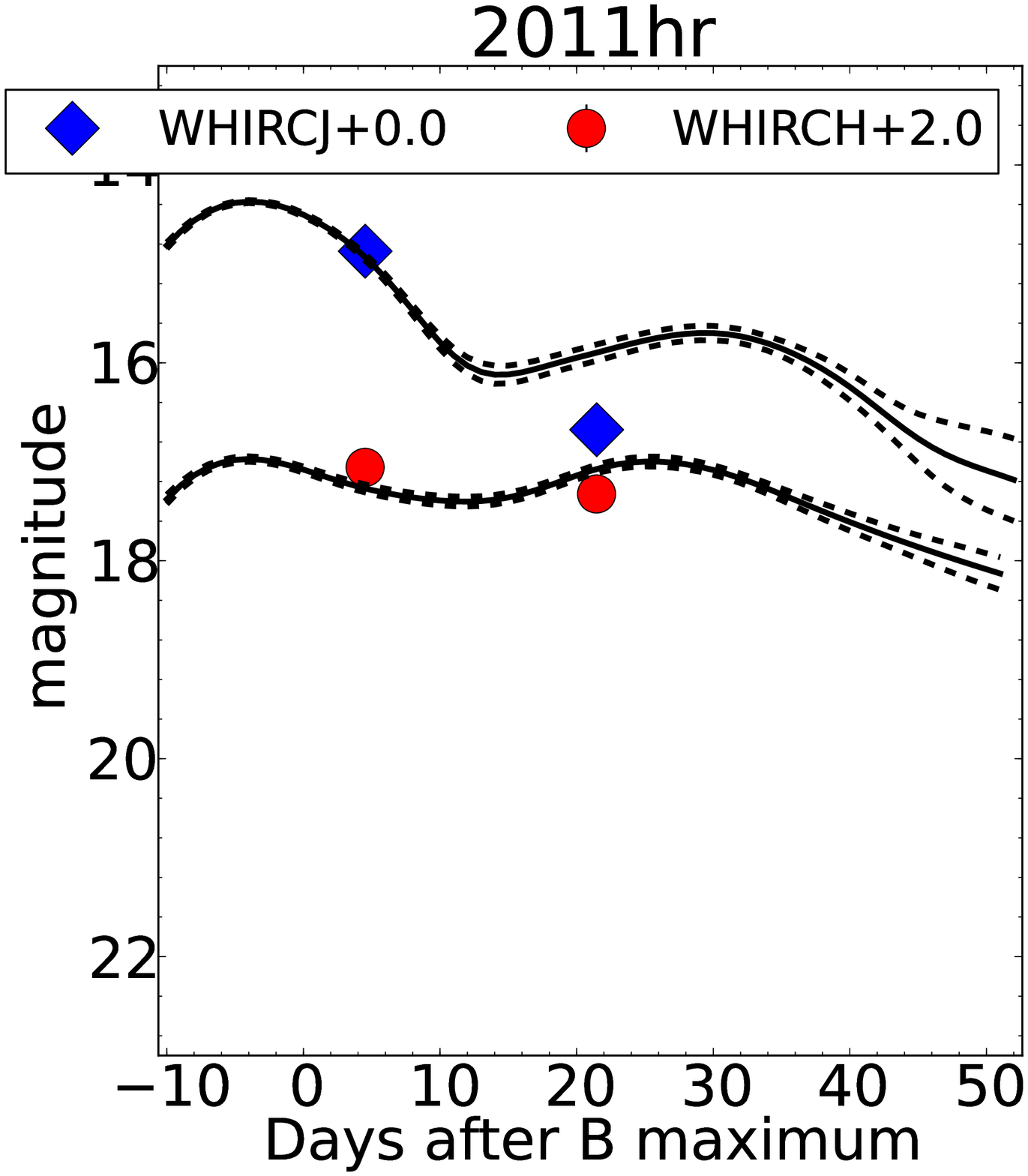}
\plotone{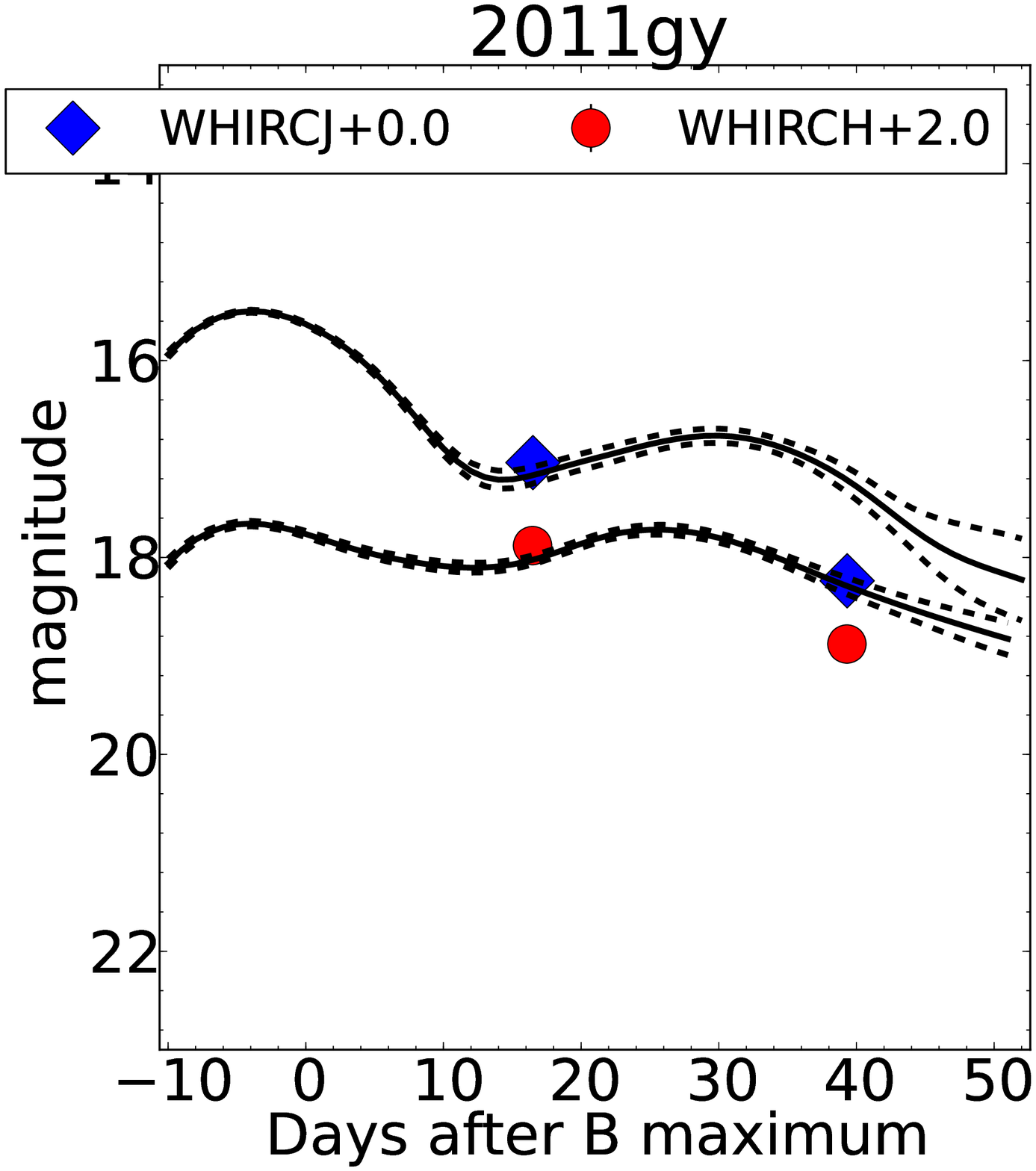}
\plotone{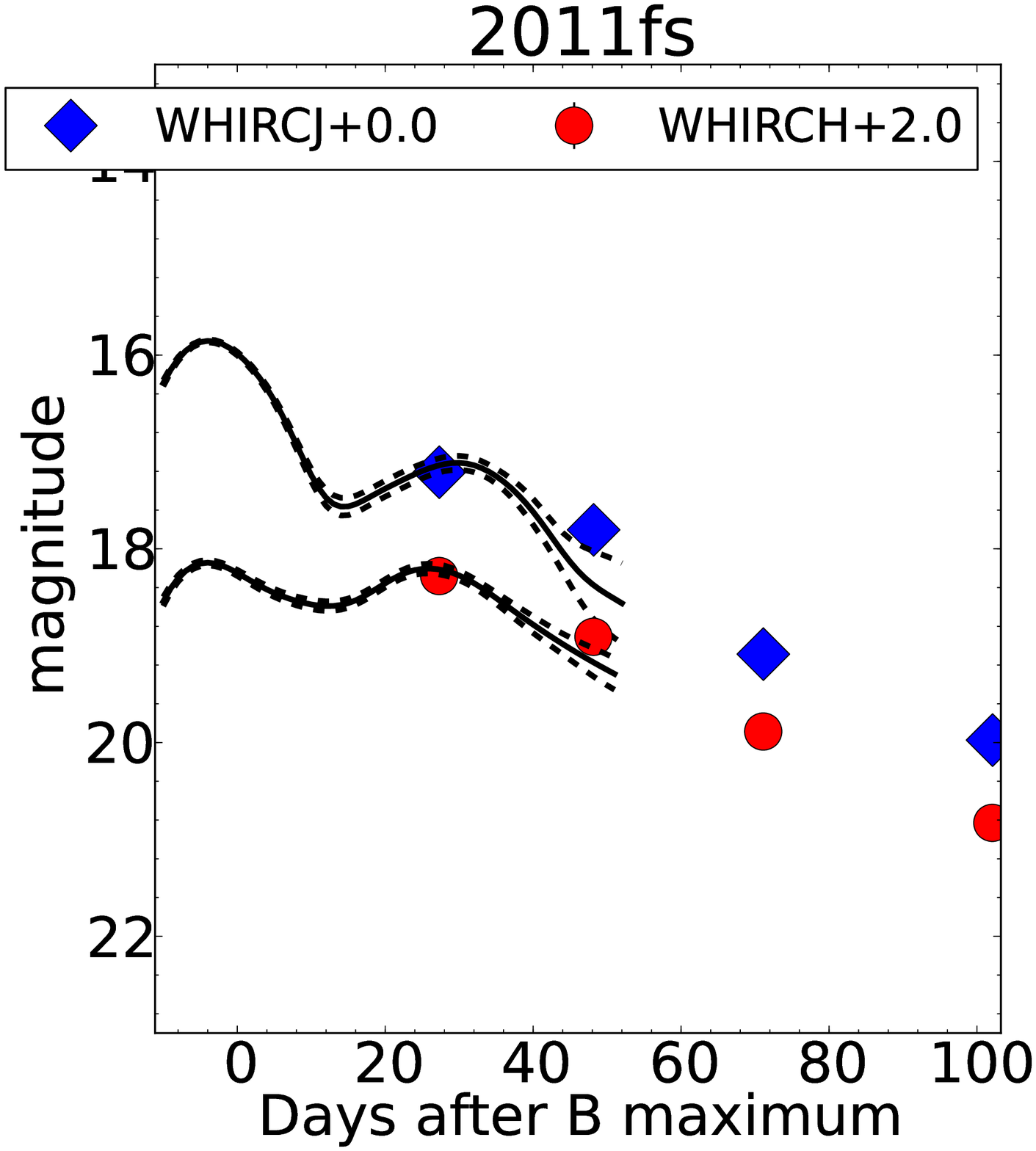} \\
\plotone{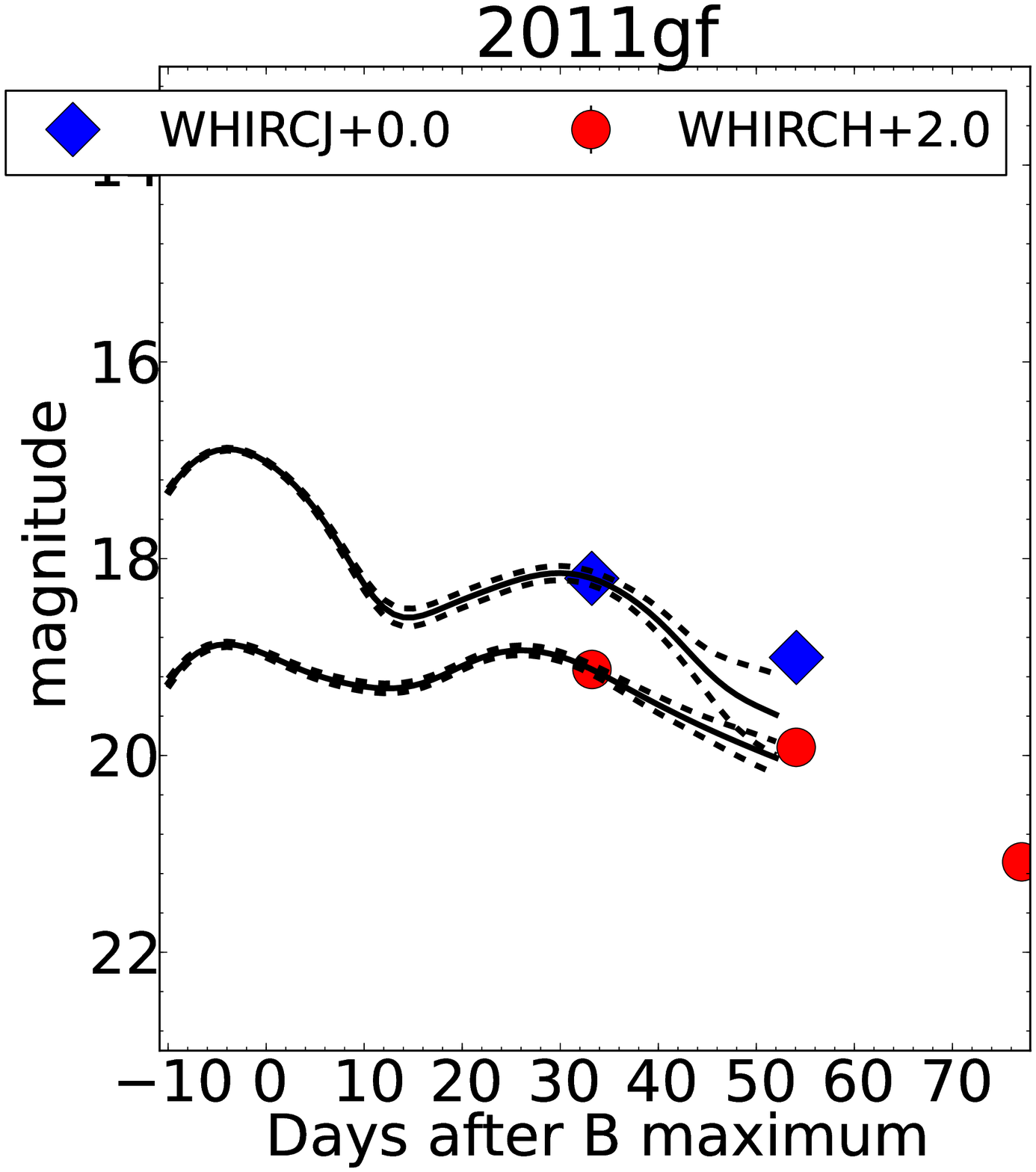}
\plotone{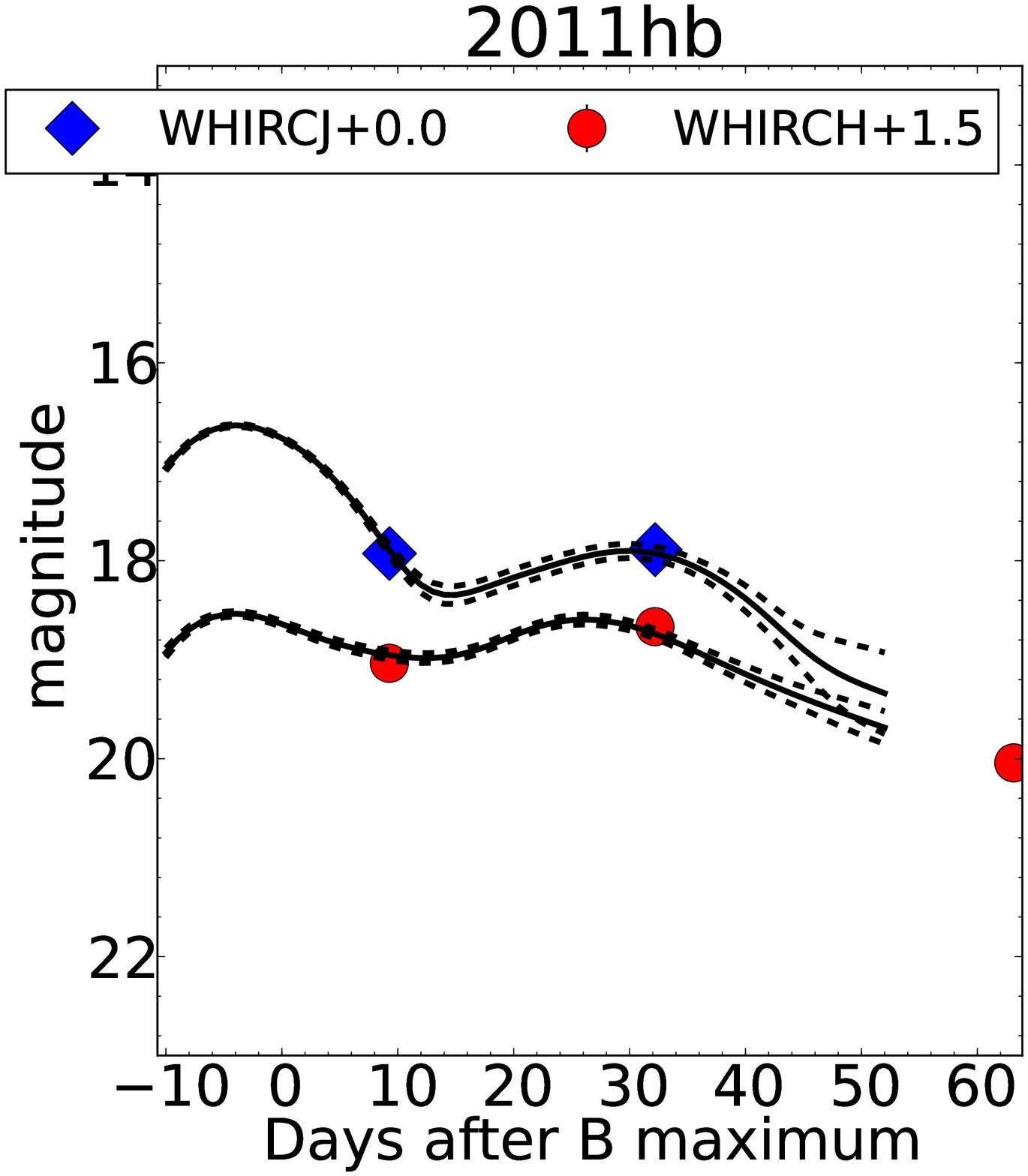}
\plotone{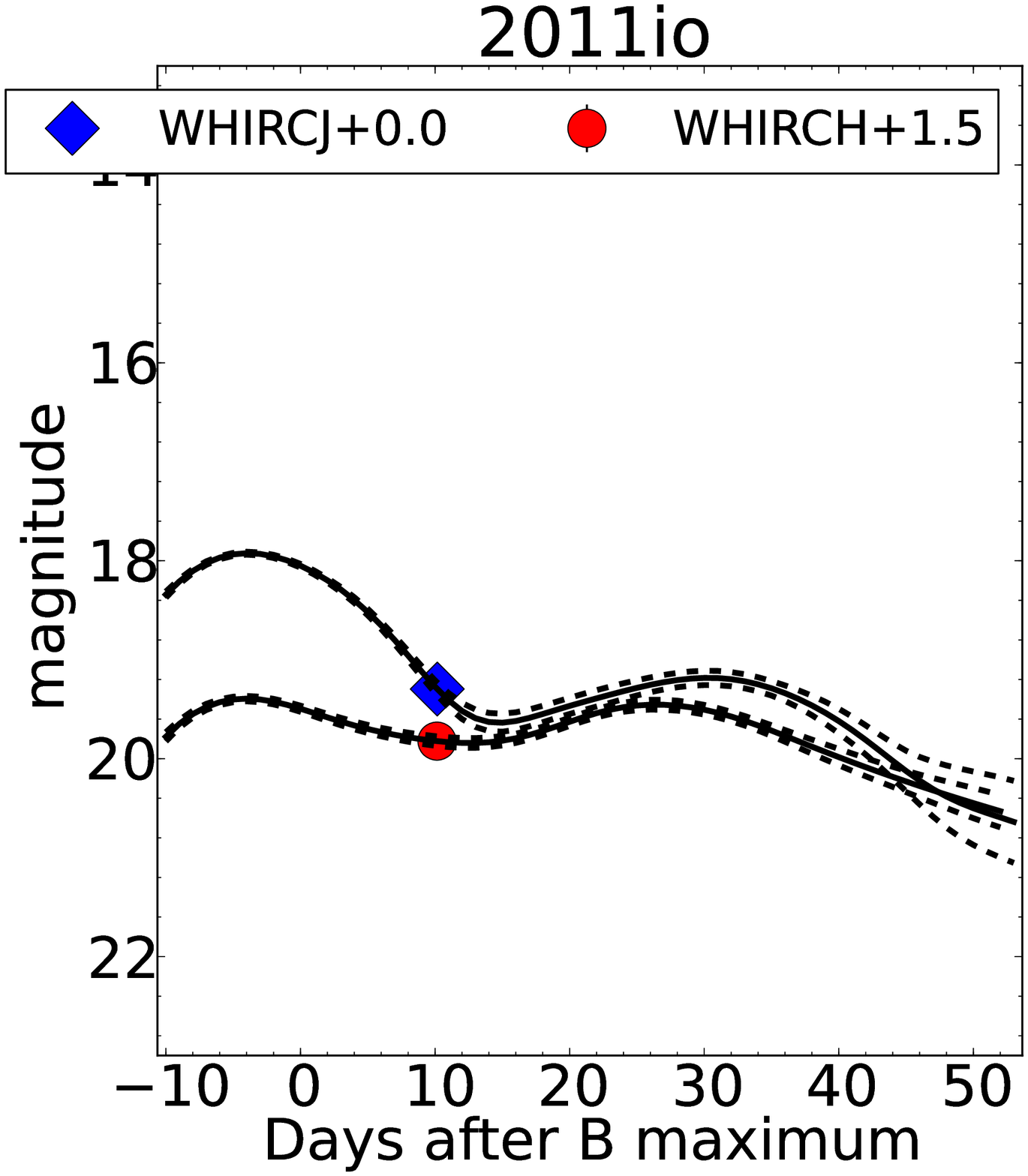} \\
\plotone{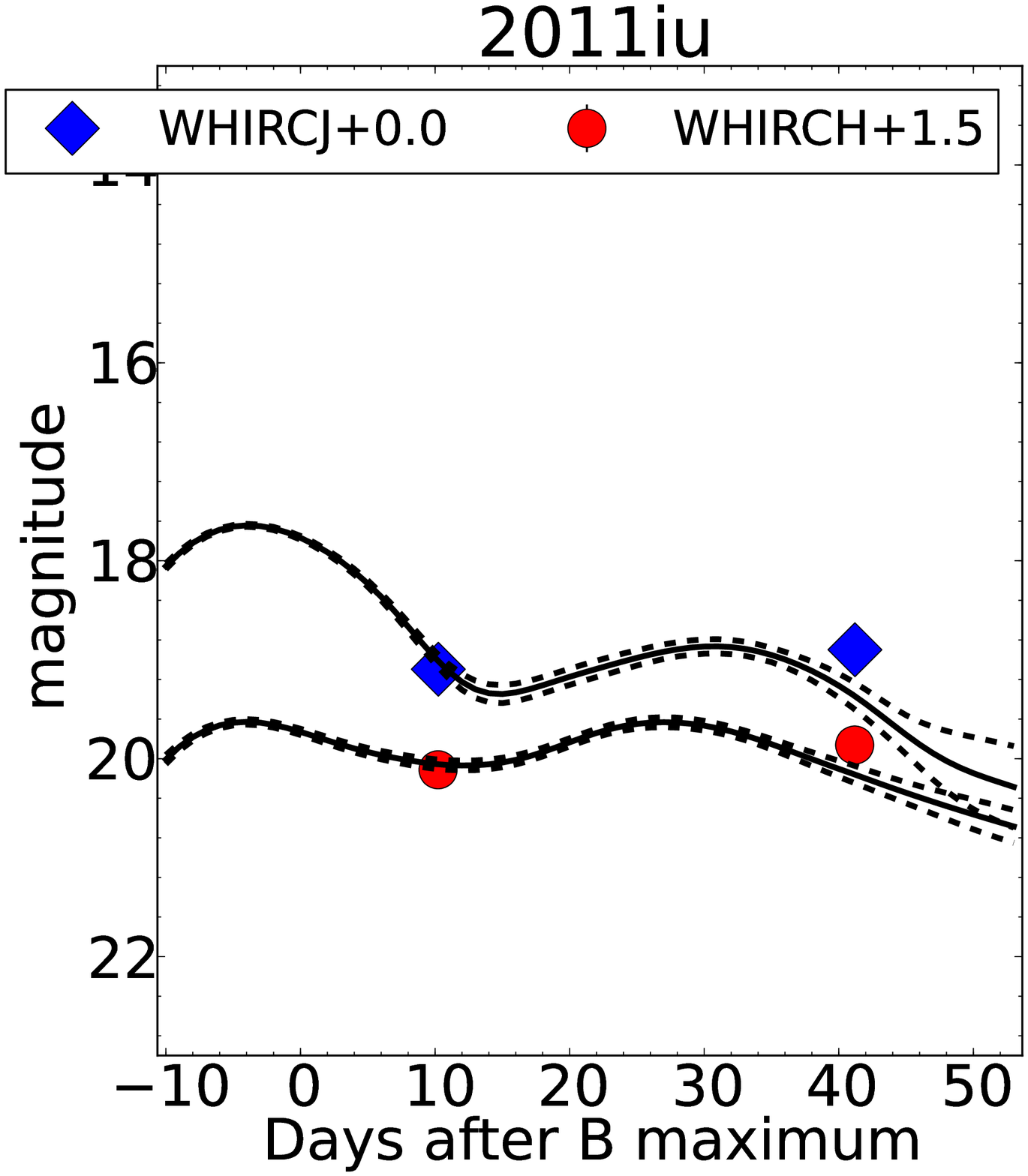}
\plotone{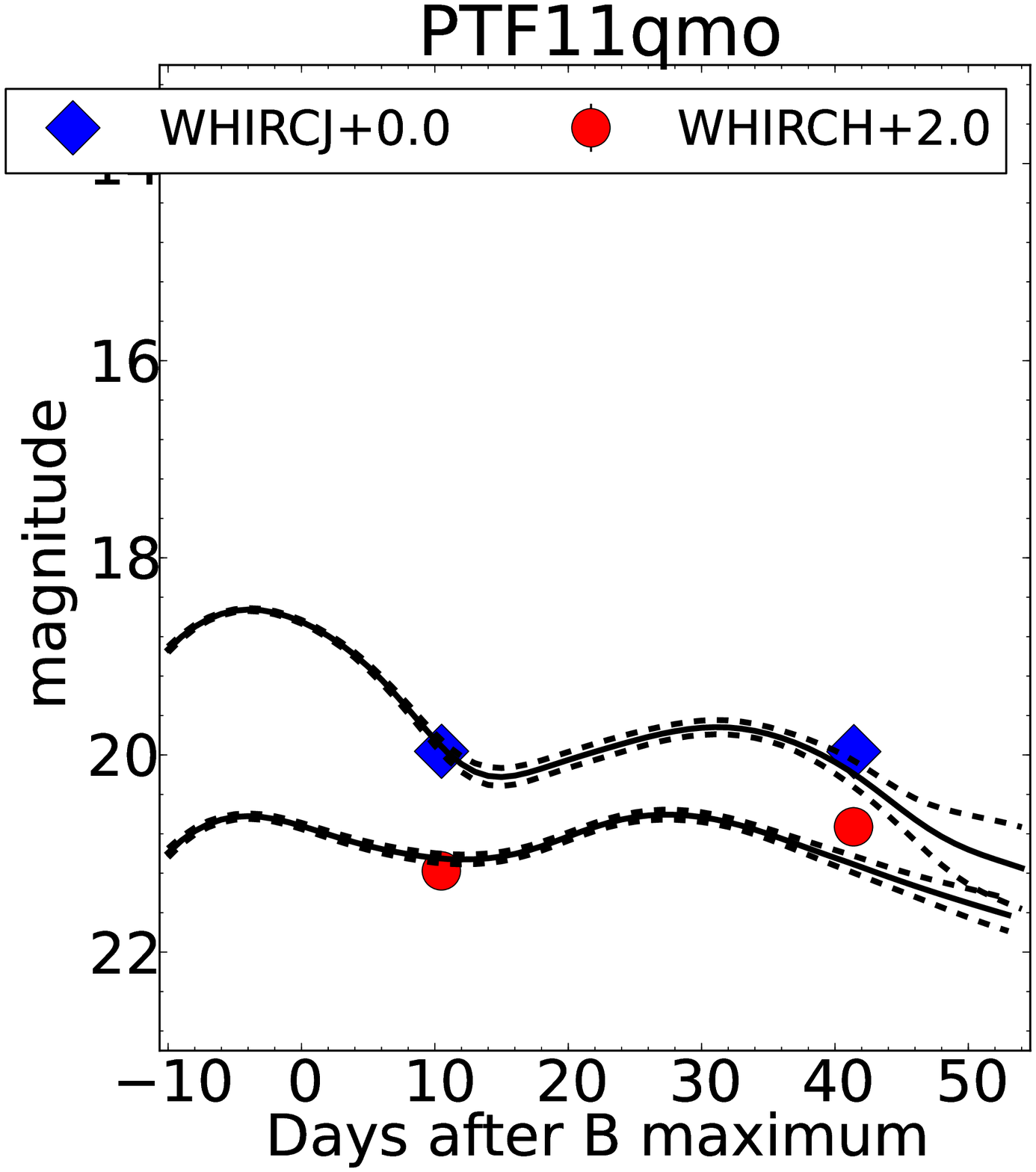}
\plotone{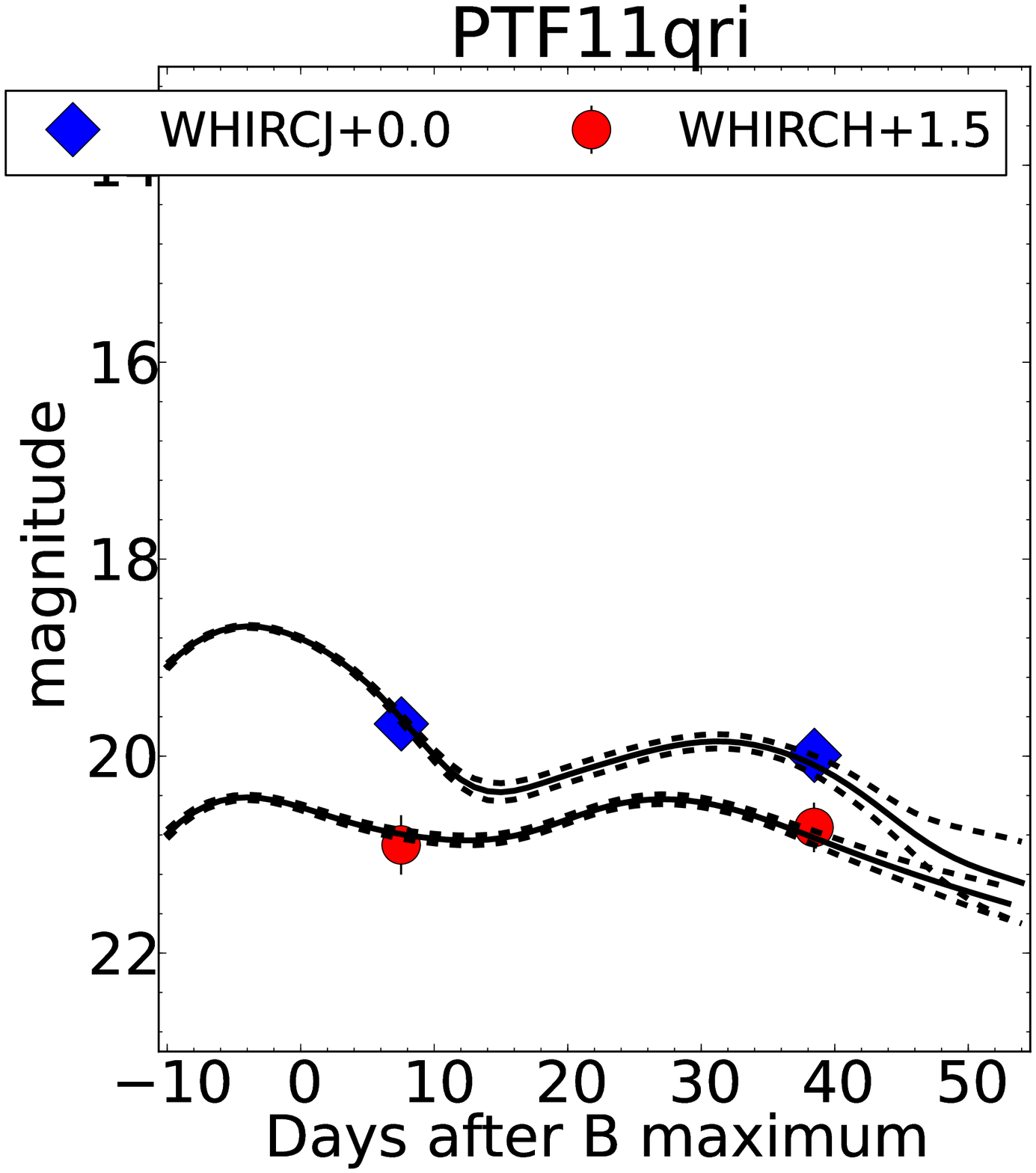} \\
\plotone{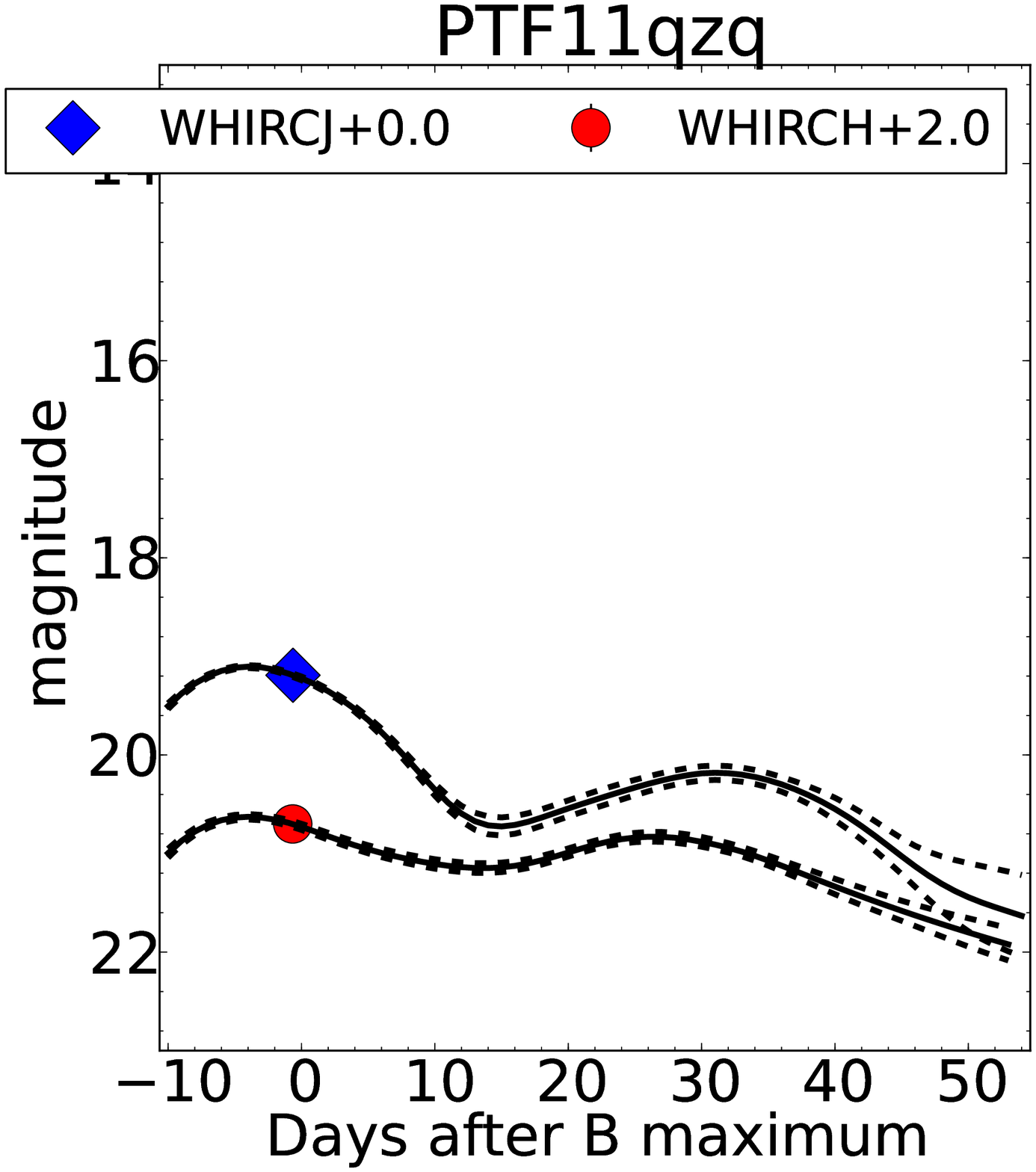}
\plotone{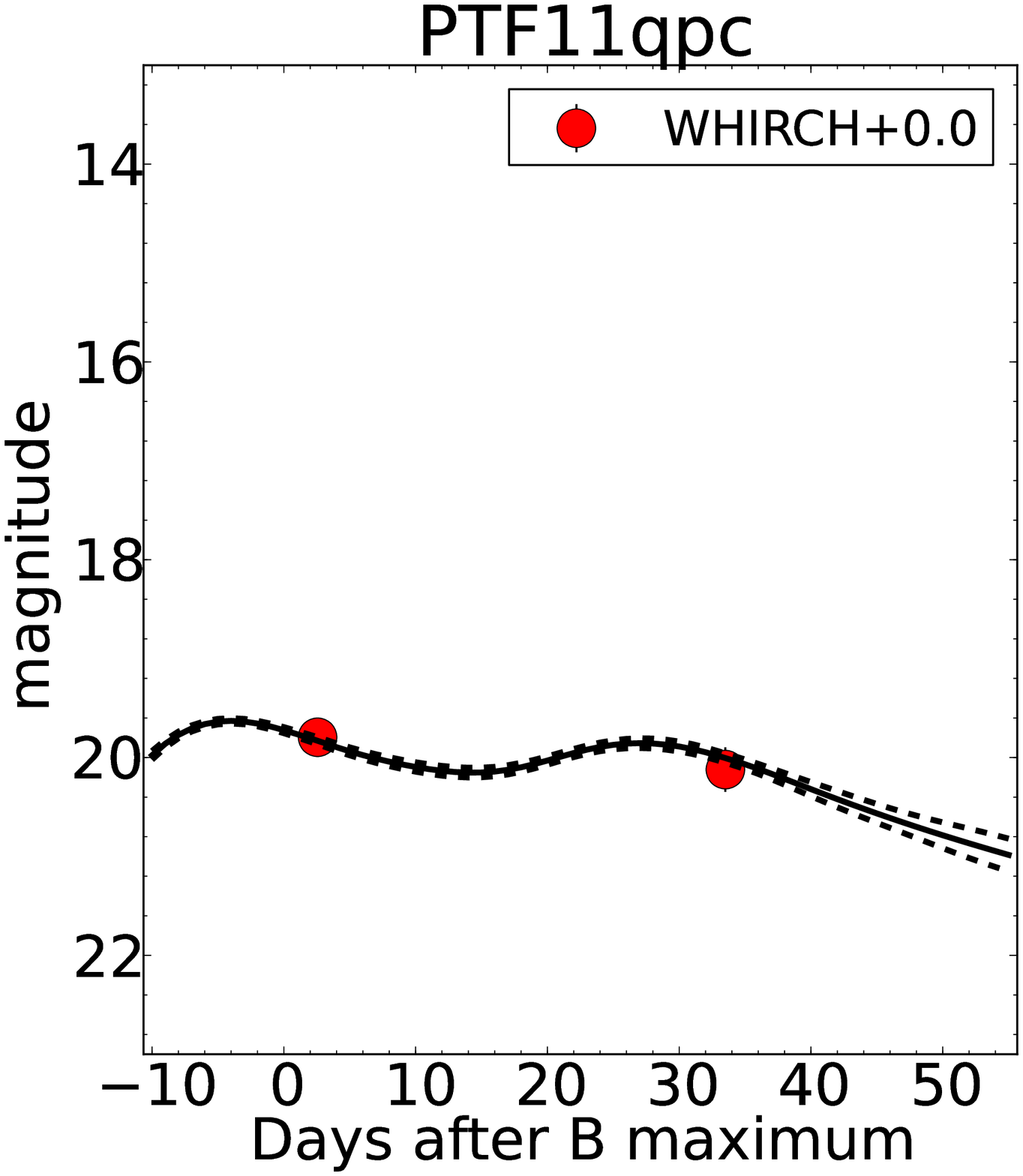}
\plotone{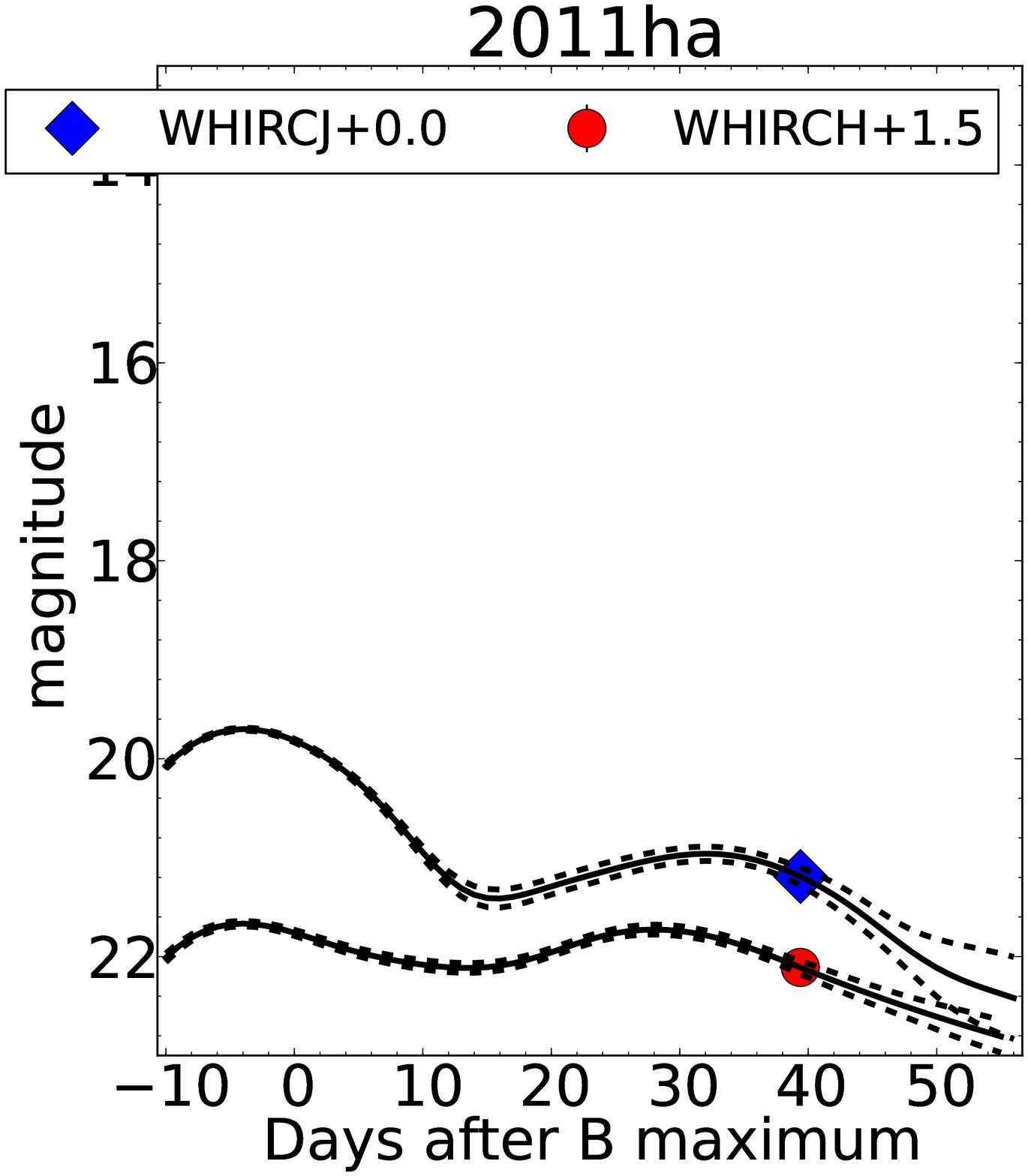}
\caption{ 
SNooPy light-curve fits for our 12 normal SNe~Ia to our $H$-band (red circle) and $J$-band (blue diamond) data.  $H$-band is offset for clarity.  
For these fits the time of maximum was fixed to the value estimated from the spectrum that was used to type the event and was reported in an ATel or CBET.  The decline-rate parameter is also fixed to $\Delta m_{15}=1.1$ making apparent magnitude the only free parameter in the fit.   SN~2011hk is not included because it was spectroscopically classified as a sub-luminous supernova similar to SN~1991bg.
}
\label{fig:lightcurvefits}
\end{figure}

For the 2011B data presented in this paper $t_{\rm max}$ is fixed to an estimate measured from the spectrum as reported in the ATels/CBETs.  This significant prior is necessary as our NIR data only have a few points per light curve (see Table \ref{table:datasummaryI}), which are not enough to independently estimate $t_{\rm max}$.
We also fix the light-curve width parameter to $\Delta m_{15}=1.1$.
This is reasonable as we have already eliminated SNe~Ia spectroscopically identified as 91bg-like from observation in our own program and from consideration when including the current literature sample.
As a result of these priors, only the peak magnitude in each filter ($JH$) is determined from fitting the light curve (see Table~\ref{table:datasummaryII}).  
The quoted peak magnitude uncertainties are then determined from least-squares fitting. 
The light-curve fits to each of the new SNe~Ia presented here are shown in Fig.~\ref{fig:lightcurvefits}.

In order to use a consistent method to compare the apparent brightness of the SNe~Ia across our entire sample we applied a similar process for the literature sample.
We use a prior on the time of maximum for the K+, CSP, and WV08 data from the SNooPy fit to the $B$-band light curve alone and fixed $\Delta m_{15}=1.1$.  
SN 2005ch is an exception as we do not have a $B$-band light curve.  We fixed the time of maximum for this SN to an estimate from the spectrum reported in \citet{CBAT8541}.
The optical light curves are not available for the BN12 data and not all SNe~Ia in this sample were reported in ATels.  We cannot estimate $t_{\rm max}$ for a fixed value of $\Delta m_{15}$ as we have done for the other samples.  
Therefore, we fixed the time of maximum and stretch to that reported for these SNe~Ia in \citet{Maguire12}.

The peak apparent magnitude for the 2011B SNe~Ia in $JH$ are listed in Table \ref{table:datasummaryII}.  
A summary of the light curve fit parameters - which includes the peak apparent magnitude - for the CSP, WV08, BN12, and the present W13 samples can be found in Table~\ref{table:HubbleFigData}.  
The W13 data is the same as that in Table~\ref{table:datasummaryII}, but we include it in Table~\ref{table:HubbleFigData} for the convenience of presenting all of the Hubble Diagram information in a single table.


\section{Results}\label{sec:results}

\subsection{Near-Infrared SN~Ia Hubble Diagram}

\begin{figure}
\epsscale{1.0}
\plotone{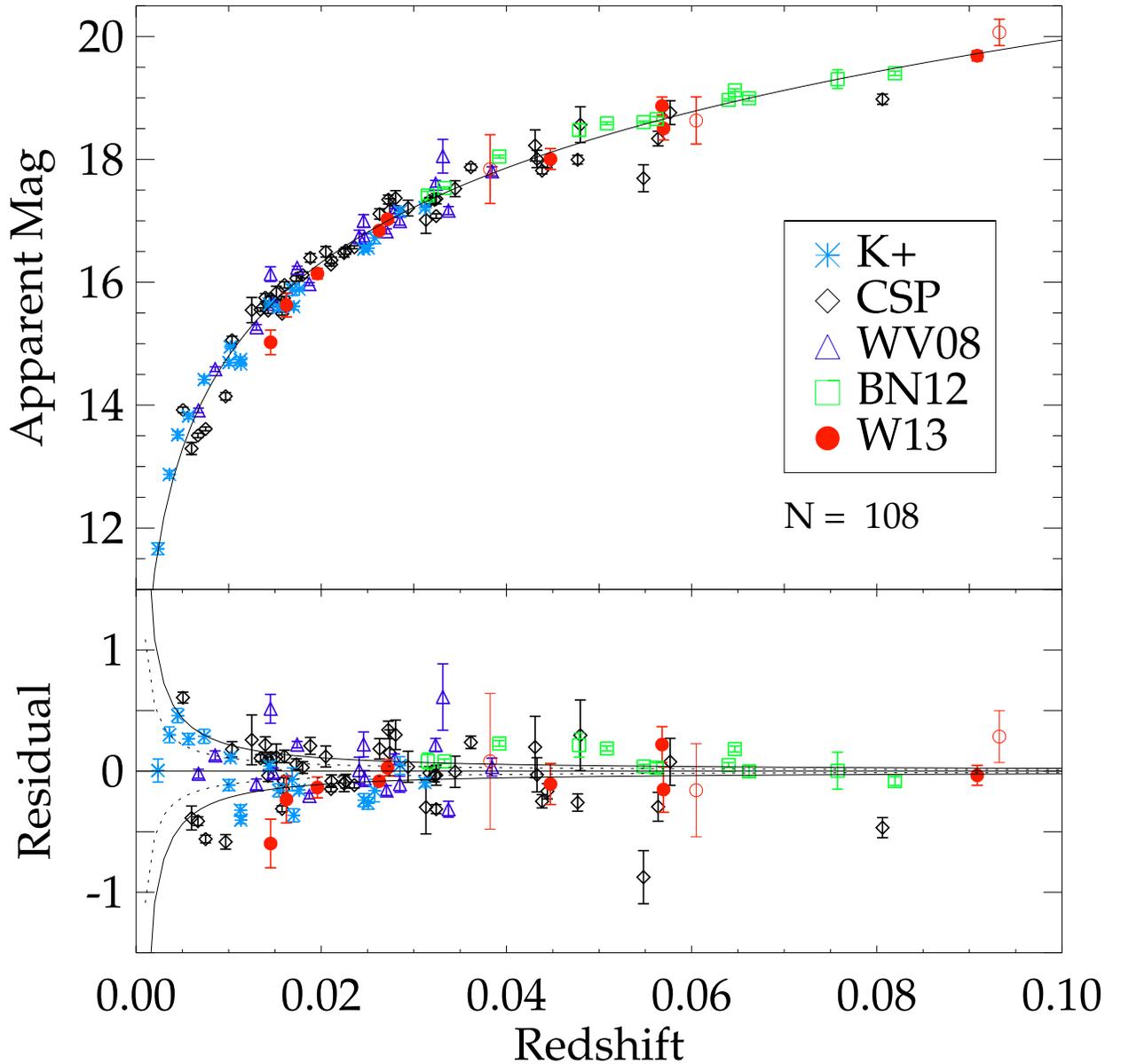}
\caption{(Top) $H$-band Hubble diagram.  The additional supernovae from this work (red circles) confirm the standard nature of SNe~Ia in $H$-band and include the two farthest SNe~Ia observed in rest-frame $H$ to date.  The open red circles indicate supernovae from our sample which have only one observation in their light curve. The model line plotted over the data is a standard flat LCDM cosmology with $\Omega_M = 0.28$.
Assuming a value of $H_0=72$~km~s$^{-1}$~Mpc$^{-1}$ 
we measure the SN~Ia $H$-band absolute magnitude from the entire sample to be \sampleMH~mag.
(Bottom) Hubble residuals (data$-$model). 
The solid (dotted) line represents the magnitude associated with a peculiar velocity uncertainty in redshift of 300~km~s$^{-1}$ (150~km~s$^{-1}$).  
Note that the largest statistical outlier from our sample, SN~2011hr, is both the lowest-redshift of our sample ($z=0.01328$) and is also spectroscopically classified as 91T-like and could be expected to be over-luminous with respect to the assumption of a fiducial SN~Ia made in our fits. 
}
\label{fig:HbandHD}
\end{figure}

An $H$-band Hubble diagram for our entire sample is presented in Fig.~\ref{fig:HbandHD}.  The recession velocities are based on Virgo infall model of \citet{Mould00} (see Table 7).  
For SNe~Ia within 3000~km~s$^{-1}$ we fix the redshifts to those summarized in \citet{Wood-Vasey08}.
The solid line in the top panel of Fig. \ref{fig:HbandHD} represents the observed apparent magnitude assuming a standard flat cosmology of $\Omega_M=0.28$ and $H_0=72$~km~s$^{-1}$~Mpc$^{-1}$ and $M_H=-18.32$~mag (see Section 5.2).  
The residuals with respect to this line are plotted in the bottom panel.  
The highest redshift outlier from CSP is SN 2005ag at $z=0.08062$.  
\citet{Folatelli10} find SN 2005ag to be a slow-decliner and therefore more luminous than a normal SN~Ia although the luminosity versus decline-rate relationship should correct for this.  They also believe that this SN was at the detection limit of LOSS such that Malmquist bias could explain its over-brightness. 

\begin{figure}
\plotone{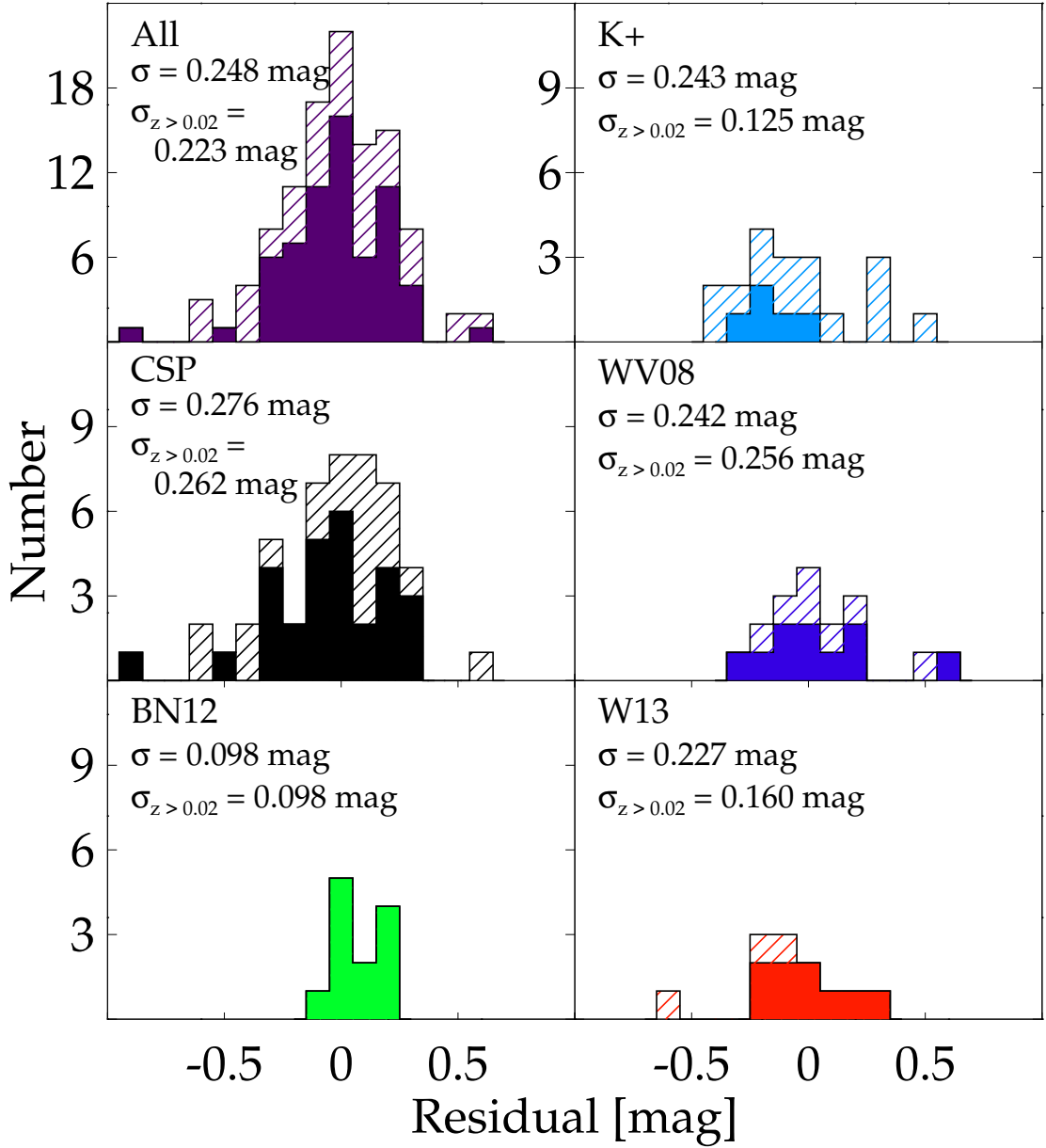}
\caption{
Distribution of the H-band residuals with respect to the global mean \sampleMH~mag. organized by survey for the entire sample (hatched) and for SN~Ia with $z > 0.02$ (solid).  Supernovae observed by WV08 and CSP are included in the WV08 sample.  The weighted standard deviation is quoted in the top right corner for the whole sample (top) and the higher redshift sub-sample (bottom).  One can clearly see the benefit of obtaining a sample in the smooth Hubble flow by the tight BN12 residual distribution and to some extent in W13.
}
\label{fig:H_residual_hist}
\end{figure}

We plot the distribution of residuals for each sub-sample in Fig.~\ref{fig:H_residual_hist} for the entire set (hatched) and for $z > 0.02$ (solid).  
The standard deviation of the residuals, $\sigma$, for each sample and for the subsample with $z > 0.02$ is given in each subpanel.
One can clearly see the smaller spread in the BN12 and W13 samples, a benefit of a higher redshift sample with reduced peculiar velocity uncertainty and photometric uncertainty. 

We find a dispersion for our W13 sample of $\sigma_H=$~\WHIRCdispersion~mag which reduces to $\sigma_H=$~\WHIRCdispersionnoHr~mag when we exclude SN~2011hr.  
SN~2011hr is 91T-like and could be expected to be over-luminous.  The dispersion is further reduced to $\sigma_H=$~\WHIRCdispersionnoHrmpts~mag if we exclude all SN with only one $H$-band observation and SN~2011hr which leaves us with 8 SNe~Ia.

\subsection{Absolute $H$-Band Magnitude of a SN~Ia}

We find the absolute $H$-band magnitude $M_H$ by calculating the weighted mean of the difference between the peak apparent magnitude and the distance modulus evaluated at the corresponding redshift assuming a standard flat $\Lambda$CDM cosmology of $\Omega_M=0.28$ and $H_0=72$~km~s$^{-1}$~Mpc$^{-1}$.  
The weight includes the additional uncertainty as a result of redshift uncertainty associated with a peculiar velocity of 150~km~s$^{-1}$ \citep{Radburn-Smith04}.
We find $M_H=$\sampleMH~mag for the entire sample.  
This value is completely degenerate with the choice of $H_0$, in the sense that a larger $H_0$ corresponds to a fainter absolute magnitude.  
So in more generality we find 
$M_H=
(\sampleMH)
+5\log_{10}{\left(H_0/(72~{\rm km}~{\rm s}^{-1}~{\rm Mpc}^{-1})\right)}$~mag.

If we analyze the measured peak $H$-band absolute magnitude separately for each sample we find:
\KMH~mag for K+, 
\CSPMH~mag for CSP, 
\WVMH~mag for WV08, 
\BNMH~mag for BN12, 
and 
\WHIRCMH~mag for W13 
(assuming the same $\Omega_M=0.28$, $H_0=72$~km~s$^{-1}$~Mpc$^{-1}$ $\Lambda$CDM cosmology).   
Note that the uncertainties quoted here are the standard error (i.e., the uncertainty in the determination of the mean)
rather than the standard deviation of the distribution around these absolute magnitudes (see Fig.~\ref{fig:H_residual_hist}).
The peak magnitude uncertainty quoted for each SN~Ia is underestimated for at least two reasons: 
(1) SNooPy only returns the statistical uncertainty from fitting and does not include any systematic uncertainties\footnote{For a list of systematic uncertainties that SNooPy fails to report see Section 4.4 of \citep{Burns11}.}   
and (2) the time of maximum is fixed such that uncertainty in the time of maximum is not propagated to the uncertainty in peak magnitude.  
As a result, we cannot calculate the uncertainty in measured peak $H$-band absolute magnitude as the uncertainty in the weighted mean.  This would underestimate the error in $M_H$.  Instead we look at the spread of the distribution of residuals as a whole to estimate the uncertainty and thus quote the standard error ($\sigma_H / \sqrt{N} $).

We consider a worst-case scenario to estimate the maximal contribution of uncertainty in $t_{\rm max}$ to the uncertainty in $M_H$ by coherently shifting $t_{\rm max}$ for the entire sample by the uncertainty in $t_{\rm max}$.  
 Excluding for a moment the W13 sample for which we do not have
an estimate of the $t_{\rm max}$ uncertainty, we find that $M_H$ shifts by 0.0017~mag indicating that 
the contribution from $t_{\rm max}$ uncertainty is negligible.  If we assume an uncertainty of 
$\pm$~2~days for the W13 sample we find a shift of 0.059~mag in the peak absolute brightness.  This means 
for our sample of 12 SN~Ia, the maximal contribution of $t_{\rm max}$ uncertainty to our estimate for $M_H$ is  
$0.059/\sqrt(12.)=0.017$~mag.''

To examine the error in $M_H$ incurred by fixing  $\Delta m_{15}$
we refit the WV08, CSP, and K+ $B$-band light curves allowing $t_{\rm max}$ and
$\Delta m_{15}$ to float.  We then use this $t_{\rm max}$ and $\Delta m_{15}$ as
fixed priors when fitting the $JH$-band light curves.
We find shifts in the measured peak $H$-band absolute magnitude of
-0.031~mag, 0.019~mag, and -0.007~mag for the CSP, K+ and WV08 samples.
These are well within our uncertainty on the measured peak apparent magnitude for
each sample.  Additionally, we find a negligible change in the $\chi^2$ per degree of freedom
between the two approaches and thus conclude that we are justified in using the simpler light-curve
model.

\section{Discussion}\label{sec:discussion}

\subsection{NIR SN~Ia as Standard Candles}

The dispersion of our W13 sample excluding SN 2011hr ($\sigma_H=\WHIRCdispersionnoHr$~mag) is comparable to that of \citet{Wood-Vasey08} who find an RMS of $0.16$~mag in $H$ and \citet{Folatelli10} who find an RMS of $0.19$~mag in $H$ when not correcting for host galaxy extinction.  Similar to our analysis, neither result makes a correction to the absolute magnitude according to the decline-rate.  

\citet{Barone-Nugent12} estimate that 1--2 points per light curve should yield a dispersion between 0.096 and 0.116 mag.  
However, these results derive from a sample with $B$-band stretch values ranging from 0.8 to 1.15.
Greater diversity in our sample is one possible explanation for our larger measured dispersion.
Our measured dispersion may be higher because most of our data is from +10 days after maximum and we have no pre-maximum data.
  Additionally, the times of maximum for our sample came from spectroscopic observations as reported in ATels and CBETs.  Spectroscopic phase determination are only precise to $\pm2$~days \citep{Blondin07} and there is potentially the equivalent of a couple of days of additional scatter from quick at-the-telescope reductions.

It is possible that the spectroscopic classification and reporting of the time of $B$-band is systematically biased in some way.  
For example, while some groups report precisely the best fit spectrum used to type the object and estimate the phase, others merely state the phase as, e.g., ``near maximum'' or ``several days after maximum.''    
We examined the implications of the extreme case of a coherent bias on $t_{\rm max}$ for the W13 estimate of $M_H$ by adding and subtracting 2 days to the prior on the time of maximum to {\em all} W13 SNe~Ia.  
We found that systematically shifting the time of maximum results in a shift of about $+0.06$~mag for +2 days and $-0.06$~mag for -2 days in $M_H$.
This coherent shift in apparent magnitude for the W13 sample is because all of our data are post-maximum light where the SNe~Ia are generally fading rather than increasing in brightness.

We also note that the SNe~Ia which comprise the W13 sample are not drawn from the faint limits of their discovery surveys.   
Therefore, Malmquist bias is unlikely to be a problem with the W13 sample.

Our analysis shows that for a set of spectroscopically normal SNe~Ia 
 using limited NIR data and a simplified light curve model which does not rely on any optical or stretch information, but rather only a prior on the time of maximum, we find an observed RMS of \WHIRCdispersionnoHr~mag that is comparable to detailed lightcurves from optical-only surveys.

\subsection{Absolute Brightness}

Our measurement of the absolute brightness for the CSP-sample is in good agreement with the literature.
Our CSP-sample results are 0.056 mag dimmer than those of \citet{Kattner12} who find $M_H=-18.432 \pm 0.017$~mag for their CSP sample of 27 well-observed NIR light curves.  
The \citet{Kattner12} analysis included a decline-rate correction.  \citet{Folatelli10} find $M_H=-18.40 \pm 0.08$ using the first set of CSP data and including no decline-rate correction, which is only 0.024 mag brighter than our analysis of the full CSP sample including up through \citet{Stritzinger11}.  

We are in slight disagreement with \citet{Barone-Nugent12} at the $1.5\sigma$ level who find $M_H=-18.30 \pm 0.04$~mag as the median absolute magnitude for their sample.  
\footnote{For this comparison we have adjusted the originally reported $M_H$ values of \citet{Barone-Nugent12} to match the common scale of $H_0=72$~km~s$^{-1}$~Mpc$^{-1}$ used in this present analysis and in \citet{Folatelli10} and \citet{Kattner12}.}

We also note that while our measurements for $M_H$ for W13, K+, CSP, and WV08 are in good agreement with each other, W13 and WV08 are in slight disagreement with the BN12 sample ($\sim 2 \sigma$) and K+ and CSP are in poor agreement with the BN12 sample ($+3\sigma$).  
Our treatment of the BN12 sample is different as we do not have access to the optical light curves.  
We did not determine $t_{{\rm max}}$ for a fixed value of stretch as we did for the other samples, but instead used the quoted $t_{{\rm max}}$ and stretch from \citet{Maguire12} as was used in \citet{Barone-Nugent12}. 
This inconsistent treatment of this sample may be part of the discrepancy with the results of other samples.  To test this, we reran the analysis on the BN12 data fixing the decline-rate parameter to $\Delta m_{15}=1.1$ and allowing the time of maximum to float.  
We found $M_H=$\BNMHfixdm~mag which is a marginal improvement in agreement. 
We speculate that additional disagreement here is caused by differences in the SNooPY~\citep{Burns11} and FLIRT\citep{Mandel09} light-curve fitters.



\section{SweetSpot: A 3-Year Survey Program with WHIRC}\label{sec:survey}

Building off the pilot program presented in this paper, we are currently engaged in a 3-year 72-night large-scale NOAO Survey (2012B-0500; PI: W.~M.~Wood-Vasey) program to image SNe~Ia in the NIR using WIYN+WHIRC. 
Our goal is to observe $\sim150$ spectroscopically confirmed nearby SNe~Ia in the NIR using WHIRC.  
We will obtain a total sample of $\sim$150 SN~Ia light curves sampled in $JH$ with 3--6 observations per light curve for the bulk of the sample and a subset of 25 SNe~Ia observed in $JH\Ks$ out to late phases ($>+30$~days) with 6--10 observations per supernova. 
If SNe~Ia are standard in the NIR with to $\sigma_H=0.1$~mag with no significant systematic bias 
then 150 SNe~Ia in the nearby Hubble 
flow will allow us to make an overall relative distance measurement to $z\sim0.05$ to 1\%.  Alternatively, we will be able to probe systematics at the few percent level, 
beyond what we are able to do today in the optical due to the significant confusion from host
galaxy dust extinction and greater dispersion in the SN~Ia optical luminosities.

We continue to rely on the hard work of several nearby supernovae surveys to discover and spectroscopically-confirm the SNe~Ia we observe.
Specifically, we follow announcements from the IAU/CBETs and ATels of supernovae discovered and/or classified by 
KAIT/LOSS \citep{Filippenko01}, 
CRTS \citep{Drake09} surveys, 
the intermediate Palomar Transient Factory\footnote{\url{http://ptf.caltech.edu/iptf/}}, 
Robotic Optical transient search experiment\footnote{\url{http://www.rotse.net}},
the Backyard Observatory Supernova Search\footnote{\url{http://bosssupernova.com}},
the Italian Supernova Search Project\footnote{\url{http://italiansupernovae.org}},
the La Silla Quest survey\footnote{\url{http://hep.yale.edu/lasillaquest}} \citep{Baltay12}, 
the CfA Supernova Group\footnote{\url{http://www.cfa.harvard.edu/supernova/SNgroup.html}}\citep{Hicken12},
the Public ESO Spectroscopic Survey of Transient Objects\footnote{\url{http://www.pessto.org/pessto/index.py}},
the Padova-Asiago Supernova Group \footnote{\url{http://graspa.oapd.inaf.it}},
and the Nearby Supernova Factory II\footnote{\url{http://snfactory.lbl.gov}}\citep{aldering02}.

We would be happy to work on collaborative efforts to analyze the SNe~Ia we are observing with those who have optical lightcurves and spectra or other near-infrared data and invite those interested to contact the first two authors (AW and MWV) to pursue such opportunities.

With this sample we will extend the SNe~Ia NIR $H$-band Hubble Diagram out to $z \sim 0.08$. This will increase the currently published sample size in this ``sweet spot'' redshift range by a factor of five.  The Carnegie Supernova Project II\footnote{\url{http://csp2.lco.cl/}} is currently engaged in a similar effort to obtain optical+NIR imaging and spectroscopy for a similar sample size in this same redshift range.

While we will obtain $6$--$10$ light curve observations for most of the SNe~Ia, we will also explore constructing the ``minimal'' $H$-band Hubble diagram.  
NIR observations are expensive to take from the ground as a result of the significant emission and absorption from the atmosphere, and expensive from space due to the cryogenic detectors often desired.  
If we could determine distances reliably with just a few NIR data points combined with an optical light curve, 
we would significantly increase the number of SN~Ia distances that could be measured for a given investment of NIR telescope time.  
We will realistically evaluate this ``minimal'' required contribution of NIR data to SN~Ia cosmology by analyzing the optical light curve with only one or two $H$-band observations near maximum and check this against the luminosity distance determined from the actual full $H$-band light curve.  
The optical light curve will give us the phase and we will measure the brightness in the near infrared.  If this approach is successful it opens the window to exploring SNe~Ia at higher redshift even given the significant cost of rest-frame NIR observations.  We will quantify the improvement of adding 1--3 NIR observations per SN~Ia and make recommendations for the most feasible and beneficial strategy for improving SN~Ia cosmology.


If modest observations of only a few restframe $H$-band points along the lightcurves of a SNe~Ia are sufficient to provide a robust and relatively precise distance measurement, then there is significant potential in supplementing future large ground-based surveys, such as the Large Synoptic Survey Telescope~\citep{LSSTScienceBook}, with space-based resources such as the James Webb Space Telescope\footnote{\url{http://www.jwst.nasa.gov/}} to obtain restframe $H$-band observations to check systematic effects in these large surveys and to independently obtain reliable NIR distances to $z>0.5$.

A newly identified systematic affecting inferred optical luminosity distances from SNe~Ia is the stellar mass of the host galaxy~\citep{Kelly10,Lampeitl10,Sullivan10,Gupta11,Childress13}.  
These analyses show that, after light-curve shape corrections,  
SNe~Ia in high-stellar-mass galaxies are found to be $0.1$~mag brighter in rest-frame $B$ than in low-stellar-mass galaxies.  
Recent work based on IFU observations of the local (1~kpc) environments of SNe~Ia \citep{Rigault13} explains this effect as a consequence of the distribution of {\em local} star-formation conditions in nearby galaxies. 
They find that a population of SNe~Ia in locally passive environments is 0.2~mag brighter than SNe~Ia in locally star-forming environments.  In higher-mass galaxies, there is an equal mix of these SNe~Ia, leading to a 0.1~mag bias, while in lower-mass galaxies ($M_\Sun<10^{9.5}$) such a bright population does not appear to exist.

The NIR photometry we will obtain of the SN host galaxies will provide both reference templates for the supernova lightcurves as well as key observations to determine stellar mass.
We will explore if these mass and environmental correlations hold in the near infrared by combining our NIR supernova observations with samples from the literature together with observations of the host galaxies.  

We will finally examine the late time color evolution of SNe~Ia in the near infrared.  SNe~Ia have a uniform optical color evolution starting around 30 days past maximum light~\citep{Lira96,Phillips99}.  
The full decay rate and color evolution from maximum light to $100$~days will provide excellent calibration of the intrinsic color and dust extinction in SNe~Ia.  
If SNe~Ia are confirmed to be standard in their NIR late-time color evolution then we can use a combined UV, optical, and NIR data set to make detailed measurements of the dust extinction in the SN~Ia host galaxies.


\section{Conclusion}\label{sec:conclusion}

We are using the WIYN 3.5m Observatory at Kitt Peak as part of an approved NOAO Survey to image nearby SN~Ia in the NIR using WHIRC.   
In this paper we have presented 13 light curves for SNe~Ia observed in 2011B as part of this program.  
Within this set we have contributed 12 new standard SNe~Ia to the current nearby NIR sample out to $z \sim 0.09$.

We have presented an updated $H$-band Hubble diagram including the latest samples from the literature.  
Considering that we have late-time sparsely sampled lightcurves and a time of maximum that is accurate to a few days, it is remarkable that we measure a dispersion of our sample to be \WHIRCdispersionnoHr~mag when excluding 91T-like SN 2011hr.  
With future semesters of observing and a larger sample of SN~Ia observed near maximum, we expect the dispersion to decrease as a result of more comprehensive temporal sampling.  
The dispersion will also improve as the optical counterparts of these SN~Ia become available and the times of maximum can be more accurately determined.


\acknowledgments
The observations presented in this paper came from NOAO time on WIYN under proposal ID 2011B-0482.
AW and MWV were supported in part by NSF AST-1028162. A.W. additionally acknowledges support from PITT PACC and the Zaccheus Daniel Foundation.
Supernova research at Rutgers University is supported in part by NSF CAREER award AST-0847157 to SWJ.
The ``Latest Supernovae'' website\footnote{\url{http://www.rochesterastronomy.org/supernova.html}} maintained by David Bishop was helpful in planning and executing these observations.
We thank Chris Burns for his significant assistance in using SNooPy.
We thank Sandhya Rao for her assistance with IRAF.
We thank the staff of KPNO and the WIYN telescope and engineering staff for their efforts that enabled these observations.
We thank the Tohono O'odham Nation for leasing their mountain to allow for astronomical research.
We thank the Aspen Center for Physics for hosting the 2010 summer workshop on ``Taking Supernova Cosmology into the Next Decade'' where the original discussions that led to the SweetSpot survey took place.

This research has made use of the NASA/IPAC Extragalactic Database (NED) which is operated by the Jet Propulsion Laboratory, California Institute of Technology, under contract with the National Aeronautics and Space Administration. 

This publication makes use of data products from the Two Micron All Sky Survey, which is a joint project of the University of Massachusetts and the Infrared Processing and Analysis Center/California Institute of Technology, funded by the National Aeronautics and Space Administration and the National Science Foundation.

{\it Facility:} \facility{WIYN}



\bibliographystyle{apj}
\bibliography{NIR}

\begin{deluxetable}{llllll}
\tabletypesize{\footnotesize}
\tablewidth{0pt}
\tablecolumns{6}
\tablecaption{SN~Ia Properties}
\tablehead{\colhead{SN}             &
           \colhead{Host Galaxy}      &
           \colhead{Spectral\tablenotemark{a}} & 
           \colhead{ATel/CBET}       &
           \colhead{Discovery Group\tablenotemark{b}/} &
           \colhead{Disc./Spec.\tablenotemark{c}} \\

	   \colhead{}  &
           \colhead{}  &
           \colhead{Subtype} & 
           \colhead{}  &
           \colhead{Individual} &
           \colhead{Reference}
}
\startdata
SN~2011hr   & NGC 2691  & 91T-like  & CBET 2901 & LOSS                  & N11, Z11b   \\
SN~2011gy   & UGC 02756 & Normal    & CBET 2871 & Z. Jin, X. Goa        & JG11, Z11a  \\ 
SN~2011hk   & NGC 0881  & 91bg-like & CBET 2892 & K. Itagaki, Y. Hirose & Na11, MB11b \\
         &           &           & ATEL 3798 & PTF                   & GY11b       \\
SN~2011fs   & UGC 11975 & Normal    & CBET 2825 & Z. Jin, X. Goa        & J11, B11    \\
SN~2011gf   & SDSS J211222.69-074913.9     & Normal    & CBET 2838 & CRTS                  & D11, M11    \\
SN~2011hb   & NGC 7674  & Normal    & CBET 2880 & CRTS                  & H11, MB11a  \\
         &           &           & ATEL 3739  & PTF                  & GY11a       \\
SN~2011io   & 2MASX J23024668+0848186  & Normal & CBET 2931 & MASTER    & BL11, F11   \\
SN~2011iu   & UGC 12809 &  Normal   & CBET 2939 & Puckett               & C11, MB11c \\
PTF11qri & LCRS B124431.1-060321     & SN Ia     & ATEL 3798 & PTF                   & GY11b \\
PTF11qmo & 2MASX J10064866-0741124  & SN Ia  & ATEL 3798 & PTF       & GY11b \\
PTF11qzq & 2MASX J07192718+5413454  & SN Ia  & ATEL 3798 & PTF       & GY11b \\
PTF11qpc & SDSS J122005.46+092418.3 &SN Ia   & ATEL 3798 & PTF       & GY11b \\
SN~2011ha   &  PGC 1375631          & Normal & CBET 2873 & MASTER       & LB11, O11
\enddata
\tablenotetext{a}{Spectral classifications according to SNID \citep{Blondin07} and PTF.  Subtypes given when provided in the original CBET or ATEL.}
\tablenotetext{b}{References/URLs: KAIT/LOSS \citep{Filippenko01}; CRTS \citep{Drake09}; PTF \url{http://www.astro.caltech.edu/ptf/}; MASTER \url{http://observ.pereplet.ru/sn\_e.html}; Puckett \url{http://www.cometwatch.com}}
\tablenotetext{c}{Reference Codes: 
 N11: \citet{CBET2901-D}; Z11b: \citet{CBET2901-S}; JG11: \citet{CBET2871-D}; 
 Z11a: \citet{CBET2871-S}; Na11: \citet{CBET2892-D}; MB11b: \citet{CBET2892-S}; 
 GY11b: \citet{ATel3798}; J11: \citet{CBET2825-D};
 B11: \citet{CBET2825-S}; D11: \citet{CBET2838-D}; M11: \citet{CBET2838-S}; 
 H11: \citet{CBET2880-D}; MB11a: \citet{CBET2880-S};
 GY11a: \citet{ATel3739}; BL11: \citet{CBET2931-D}; F11: \citet{CBET2931-S}; 
 C11: \citet{CBET2939-D}; MB11c: \citet{CBET2939-S}; 
 LB11: \citet{CBET2873-D}; O11: \citet{CBET2873-S}}
\label{table:datacitation}
\end{deluxetable}

\begin{deluxetable}{lrrllcl}
\tabletypesize{\footnotesize}
\tablewidth{0pt}
\tablecolumns{7}
\tablecaption{SN~Ia Sample Summary I}
\tablehead{\colhead{Name}             &
           \colhead{RA(J2000)}        &
           \colhead{Dec(J2000)}       &
           \colhead{$t_{\rm max}$\tablenotemark{a}} & 
           \colhead{$z_{\rm helio}$} &
           \colhead{$z$ from } &
           \colhead{Redshift Citation} \\
    
           \colhead{} &
           \colhead{} &
           \colhead{} &
           \colhead{} & 
           \colhead{} &
           \colhead{Host or SN} &
           \colhead{}    
}
\startdata
SN~2011hr    & 08:54:46.03 & +39:32:16.1 & 55883 & 0.01328 & Host & \citet{deVaucouleurs91}\tablenotemark{b}\\ 
SN~2011gy    & 03:29:35.30 & +40:52:02.9 & 55865 & 0.01688 & Host & \citet{Falco99}\tablenotemark{b} \\
SN~2011hk    & 02:18:45.84 & -06:38:30.3 & ...   & 0.01756 & Host & \citet{Bottinelli93}\tablenotemark{b} \\
SN~2011fs    & 22:17:19.52 & +35:34:50.0 & 55833 & 0.02091 & Host & \citet{Fisher95}\tablenotemark{b} \\
SN~2011gf    & 21:12:24.27 & -07:48:52.0 & 55827 & 0.02766 & Host & \citet{SDSSDR1}\tablenotemark{b} \\
SN~2011hb    & 23:27:55.52 & +08:46:45.0 & 55872 & 0.02892 & Host & \citet{Nishiura00}\tablenotemark{b} \\
SN~2011io    & 23:02:47.59 & +08:48:09.8 & 55894 & 0.04    & SN   & \citet{CBET2931-S} \\
SN~2011iu    & 23:51:02.27 & +46:43:21.7 & 55894 & 0.04598 & Host & \citet{Bottinelli93}\tablenotemark{b} \\
PTF11qri     & 12:47:06.28 & -06:19:49.7 & 55897 & 0.055   & SN   & \citet{ATel3798} \\
PTF11qmo     & 10:06:49.76 & -07:41:12.3 & 55894 & 0.05523 & Host & \citet{Jones09}\tablenotemark{b} \\
PTF11qzq     & 07:19:27.24 & +54:13:48.0 & 55905 & 0.06    & SN   & \citet{ATel3798} \\
PTF11qpc     & 12:20:05.47 & +09:24:12.1 & 55902 & 0.08902 & Host & \citet{SDSSDR3}\tablenotemark{b} \\
SN~2011ha    & 03:57:40.87 & +10:09:55.2 & 55842 & 0.094   & SN   & \citet{CBET2873-S} 
\enddata
\tablenotetext{a}{Time of maximum in the $B$-band according to SNID/PTF reported in CBET/ATel.}
\tablenotetext{b}{Heliocentric redshifts citations via NASA/IPAC Extragalactic Database (NED) \url{http://ned.ipac.caltech.edu/}.}
\label{table:datasummaryI}
\end{deluxetable}

\begin{deluxetable}{llccllll}
\tabletypesize{\footnotesize}
\tablewidth{0pt}
\tablecolumns{8}
\tablecaption{SN~Ia Sample Summary II}
\tablehead{\colhead{Name}             & 
           \colhead{$z_{\rm CMB+VIRGO}$\tablenotemark{a}}    &
           \colhead{nobs$_J$}   &
           \colhead{nobs$_H$}   &
           \colhead{$m_{J{\rm ,max}}$}  &
           \colhead{$\sigma(m_{J{\rm,max}})$\tablenotemark{b}}  &
           \colhead{$m_{H{\rm ,max}}$} &
           \colhead{$\sigma( m_{H{\rm,max}})$\tablenotemark{b}}

\\
  \colhead{}              &
  \colhead{}              &
  \colhead{}              &
  \colhead{}              &
  \colhead{[mag]}         &
  \colhead{[mag]}         &
  \colhead{[mag]}         &
  \colhead{[mag]}          
}
\startdata
SN~2011hr    & 0.01453 &  2 & 2 & 14.352 & 0.220 & 15.022 & 0.200 \\ 
SN~2011gy    & 0.01623 &  2 & 2 & 15.300 & 0.285 & 15.630 & 0.194 \\ 
SN~2011hk    & 0.01625 &  2 & 2 & ... & ... & ... & ... \\
SN~2011fs    & 0.01958 &  4 & 4 & 15.727 & 0.123 & 16.141 & 0.085 \\ 
SN~2011gf    & 0.02626 &  2 & 3 & 16.814 & 0.020 & 16.841 & 0.010 \\ 
SN~2011hb    & 0.02715 &  2 & 3 & 16.623 & 0.105 & 17.026 & 0.068 \\ 
SN~2011io    & 0.04 $\pm$ 0.01 & 1 & 1 & 17.817 & 0.558 & 17.841 & 0.560 \\ 
SN~2011iu    & 0.04475 &  2 & 2 &  17.640 & 0.232 & 18.005 & 0.169 \\ 
PTF11qri     & 0.057 $\pm$  0.001 & 2 & 2 & 18.769 & 0.122 & 18.689 & 0.147 \\ 
PTF11qmo     & 0.05696 &  2 & 2 & 18.621 & 0.265 & 18.503 & 0.188  \\ 
PTF11qzq     & 0.06 $\pm$  0.01 & 1 & 1 & 19.122 & 0.377 & 18.634 & 0.383 \\ 
PTF11qpc     & 0.09084 & 0 & 2 & \nodata & \nodata & 19.687 & 0.082  \\ 
SN~2011ha    & 0.093 $\pm$ 0.001 & 1 & 1 & 19.520 & 0.152 & 20.067 &  0.214 
\enddata
\label{table:datasummaryII}
\tablenotetext{a}{We follow \citet{Mould00} to correct for the Virgo cluster and transform to the CMB using \citet{Karachentsev96} and \citet{Fixsen96}.}
\tablenotetext{b}{Error includes photometric and redshift uncertainty as well as uncertainty from the template used to fit the data.}
\end{deluxetable}

\begin{deluxetable}{llcc}
\tabletypesize{\footnotesize}
\tablewidth{0pt}
\tablecolumns{4}
\tablecaption{Photometric Calibration Terms}
\tablehead{
  \colhead{Filter}     &
  \colhead{zeropoint}  &
  \colhead{$k$}          &
  \colhead{$c$} 
\\
  \colhead{}              &
  \colhead{[mag]}         &
  \colhead{[mag/airmass]} &
  \colhead{}
}
\startdata
$J$ & 27.041 $\pm$  0.012 & $-$0.051 $\pm$  0.020  & $+$0.062 $\pm$ 0.035 \\
$H$ & 27.140 $\pm$  0.014 & $-$0.066 $\pm$  0.030  & $-$0.186 $\pm$ 0.043 
\enddata
\label{table:transform}
\end{deluxetable}

\begin{deluxetable}{llllcccccccc}
\rotate
\tabletypesize{\footnotesize}
\tablewidth{0pt}
\tablecolumns{12}
\tablecaption{2MASS Calibration Stars}
\tablehead{
  \colhead{} &
  \colhead{} &
  \multicolumn{4}{c}{WHIRC Natural System} &
  \multicolumn{6}{c}{2MASS Catalog Magnitudes}
\\
  \colhead{2MASS ID}        &
  \colhead{SN Field}  &
  \colhead{$m_J$}        &
  \colhead{$\sigma(m_J)$}  &
  \colhead{$m_H$} &
  \colhead{$\sigma(m_H)$} &
  \colhead{$m_J$}        &
  \colhead{$\sigma(m_J)$}  &
  \colhead{$m_H$} &
  \colhead{$\sigma(m_H)$} &
  \colhead{$m_{\Ks}$} &
  \colhead{$\sigma(m_{\Ks})$} 
\\
  \colhead{} &
  \colhead{} &
  \multicolumn{2}{c}{[mag]} &
  \multicolumn{2}{c}{[mag]} &
  \multicolumn{2}{c}{[mag]} &
  \multicolumn{2}{c}{[mag]} &
  \multicolumn{2}{c}{[mag]}  
}
\startdata
2MASS 02184937$-$0637528 & SN~2011hk &          15.162 &      0.021 &     14.384 &      0.029 &  15.022 &    0.045 &   14.408 &    0.047 &   14.257  &   0.059 \\
2MASS 03293834$+$4051347 & SN~2011gy &          16.640 &      0.025 &     15.883 &      0.040 &  16.565 &    0.102 &   15.827 &    0.122 &   15.450  &   0.123 \\
2MASS 03573901$+$1009372 & SN~2011ha &          14.570 &      0.015 &     14.119 &      0.021 &  14.592 &    0.033 &   14.117 &    0.041 &   13.925  &   0.051 \\
2MASS 07192306$+$5414071 &  PTF11qzq &          16.788 &      0.022 &     16.060 &      0.042 &  16.725 &    0.127 &   15.915 &    0.145 &   \nodata &  \nodata \\
2MASS 08544039$+$3933230 & SN~2011hr &          15.526 &      0.015 &     14.923 &      0.023 &  15.587 &    0.054 &   14.903 &    0.070 &   14.738  &   0.085 \\
2MASS 10064485$-$0740334 &  PTF11qmo &          16.325 &      0.022 &     15.570 &      0.038 &  16.376 &    0.109 &   15.583 &    0.099 &   15.429  &   0.221 \\
2MASS 12200392$+$0925144 &  PTF11qpc &           ...   &       ...  &     13.728 &      0.021 &  14.482 &    0.036 &   13.779 &    0.043 &   13.526  &   0.050 \\
2MASS 12470715$-$0620106 &  PTF11qri &          15.019 &      0.019 &     14.770 &      0.030 &  15.017 &    0.029 &   14.673 &    0.060 &   14.757  &   0.096 \\
2MASS 21122081$-$0748443 & SN~2011gf &          15.131 &      0.020 &     14.317 &      0.029 &  15.171 &    0.052 &   14.389 &    0.062 &   14.280  &   0.068 \\
2MASS 22172193$+$3533349 & SN~2011fs &          15.708 &      0.020 &     15.423 &      0.032 &  15.686 &    0.056 &   15.517 &    0.113 &   15.653  &   0.244 \\
2MASS 23024227$+$0848225 & SN~2011io &          15.875 &      0.019 &     15.529 &      0.030 &  15.732 &    0.070 &   15.163 &    0.090 &   14.966  &   0.128 \\
2MASS 23275179$+$0846392 & SN~2011hb &          15.745 &      0.024 &     15.021 &      0.037 &  15.684 &    0.067 &   14.978 &    0.099 &   14.838  &   0.097 \\
2MASS 23505996$+$4643586 & SN~2011iu &          15.389 &      0.018 &     14.760 &      0.026 &  15.379 &    0.055 &   14.830 &    0.057 &   14.461  &   0.071 
\enddata
\label{table:calibrationstars}
\end{deluxetable}

\begin{deluxetable}{lllcccc}
\tabletypesize{\footnotesize}
\tablewidth{0pt}
\tablecolumns{7}
\tablecaption{SN~Ia Light Curves}
\tablehead{
  \colhead{Name}        &
  \colhead{Date}        &
  \colhead{Filter}      &
  \colhead{$m$}\tablenotemark{a} &
  \colhead{$\sigma({m})$} &
  \colhead{$\Delta_{m}^{\rm Kcorr~}$\tablenotemark{b}}
\\
  \colhead{} &
  \colhead{MJD} &
  \colhead{} &
  \colhead{[mag]} & 
  \colhead{[mag]} & 
  \colhead{[mag]} & 
}
\startdata
SN~2011hr & 55887.52 & $J$ & 14.872 & 0.024 & -0.042 \\ 
SN~2011hr & 55904.47 & $J$ & 16.676 & 0.037 & 0.023 \\ 
SN~2011hr & 55887.52 & $H$ & 15.056 & 0.036 & -0.073 \\
SN~2011hr & 55904.46 & $H$ & 15.325 & 0.037 & -0.114 \\
SN~2011gy & 55881.50 & $J$ & 17.036 & 0.040 & -0.009 \\ 
SN~2011gy & 55904.32 & $J$ & 18.237 & 0.051 & -0.017 \\ 
SN~2011gy & 55881.47 & $H$ & 15.879 & 0.045 & -0.089 \\
SN~2011gy & 55904.30 & $H$ & 16.879 & 0.057 & -0.062 \\
SN~2011hk & 55881.36 & $J$ & 17.572 & 0.024 & ... \\
SN~2011hk & 55904.28 & $J$ & 19.671 & 0.071 & ... \\
SN~2011hk & 55881.34 & $H$ & 17.027 & 0.033 & ... \\ 
SN~2011hk & 55904.26 & $H$ & 18.415 & 0.057 & ... \\ 
SN~2011fs & 55860.31 & $J$ & 17.209 & 0.038 & -0.016 \\ 
SN~2011fs & 55881.17 & $J$ & 17.804 & 0.029 & -0.025 \\ 
SN~2011fs & 55904.12 & $J$ & 19.087 & 0.045 & -0.016 \\ 
SN~2011fs & 55935.11 & $J$ & 19.975 & 0.185 & 0.000 \\ 
SN~2011fs & 55860.30 & $H$ & 16.281 & 0.040 & -0.072 \\ 
SN~2011fs & 55881.16 & $H$ & 16.908 & 0.035 & -0.063 \\ 
SN~2011fs & 55904.10 & $H$ & 17.886 & 0.044 & -0.063 \\ 
SN~2011fs & 55935.08 & $H$ & 18.829 & 0.135 & 0.000 \\ 
SN~2011gf & 55860.22 & $J$ & 18.200 & 0.044 & -0.065 \\ 
SN~2011gf & 55881.08 & $J$ & 19.004 & 0.046 & -0.066 \\ 
SN~2011gf & 55860.23 & $H$ & 17.126 & 0.045 & -0.042 \\ 
SN~2011gf & 55881.07 & $H$ & 17.917 & 0.052 & -0.054 \\ 
SN~2011gf & 55904.07 & $H$ & 19.081 & 0.188 & 0.000 \\ 
SN~2011hb & 55881.29 & $J$ & 17.927 & 0.035 & -0.083 \\ 
SN~2011hb & 55904.20 & $J$ & 17.888 & 0.025 & -0.072 \\ 
SN~2011hb & 55881.28 & $H$ & 17.536 & 0.043 & -0.032 \\ 
SN~2011hb & 55904.18 & $H$ & 17.166 & 0.038 & -0.034 \\ 
SN~2011hb & 55935.14 & $H$ & 18.542 & 0.111 & -0.048 \\ 
SN~2011io & 55904.16 & $J$ & 19.172 & 0.058 & -0.124 \\ 
SN~2011io & 55904.14 & $H$ & 18.343 & 0.055 & 0.020 \\ 
SN~2011iu & 55904.24 & $J$ & 19.096 & 0.033 & -0.141 \\ 
SN~2011iu & 55935.20 & $J$ & 18.899 & 0.114 & -0.198 \\ 
SN~2011iu & 55904.22 & $H$ & 18.612 & 0.038 & 0.047 \\ 
SN~2011iu & 55935.18 & $H$ & 18.362 & 0.104 & -0.060 \\ 
PTF11qri & 55904.54 & $J$ & 19.672 & 0.129 & -0.147 \\ 
PTF11qri & 55935.47 & $J$ & 19.992 & 0.146 & -0.268 \\ 
PTF11qri & 55904.52 & $H$ & 19.402 & 0.301 & 0.027 \\ 
PTF11qri & 55935.45 & $H$ & 19.224 & 0.251 & -0.039 \\ 
PTF11qmo & 55904.50 & $J$ & 19.963 & 0.075 & -0.175 \\ 
PTF11qmo & 55935.42 & $J$ & 19.966 & 0.163 & -0.275 \\ 
PTF11qmo & 55904.49 & $H$ & 19.176 & 0.068 & 0.083 \\ 
PTF11qmo & 55935.39 & $H$ & 18.729 & 0.099 & -0.058 \\ 
PTF11qzq & 55904.36 & $J$ & 19.056 & 0.043 & -0.136 \\ 
PTF11qzq & 55904.34 & $H$ & 18.635 & 0.078 & -0.065 \\ 
PTF11qpc & 55904.56 & $H$ & 19.795 & 0.108 & -0.079 \\ 
PTF11qpc & 55935.50 & $H$ & 20.122 & 0.225 & 0.126 \\ 
SN~2011ha & 55881.40 & $J$ & 20.434 & 0.130 & -0.756 \\ 
SN~2011ha & 55881.38 & $H$ & 20.627 & 0.191 & 0.018  

\enddata
\tablenotetext{a}{Magnitudes reported in the WHIRC natural system, which is referenced to 2MASS at $(m_J^{\rm 2MASS}-m_H^{\rm 2MASS})=0.5$~mag.}
\tablenotetext{b}{K-correction as calculated by SNooPY \citep{Burns11}.  Subtract K-correction value (column 6) from reported natural-system magnitude (column 4) to yield K-corrected magnitude in the CSP system \citep{Stritzinger11}.}
\label{table:lightcurves}
\end{deluxetable}

\begin{deluxetable}{llllllll}
\tabletypesize{\footnotesize}
\tablewidth{0pt}
\tablecolumns{8}
\tablecaption{$H$-band Maximum Apparent Magnitude for Current Sample}
\tablehead{
  \colhead{Name}        &
  \colhead{$t_{\rm max}$\tablenotemark{a}}        &
  \colhead{$z_{\rm CMB}$}      &
  \colhead{$\sigma( z_{\rm CMB})$}      &
  \colhead{$m_{H{\rm ,max}}$} &
  \colhead{$\sigma( m_{H{\rm,max}})$} &
  \colhead{Reference\tablenotemark{b}} &
  \colhead{Sample\tablenotemark{c}}
\\
  \colhead{} &
  \colhead{} &
  \colhead{} &
  \colhead{} &
  \colhead{[mag]} &
  \colhead{[mag]} &
  \colhead{} &
  \colhead{} 
}
\startdata
SN~1998bu &    50953.4 &     0.0024 & 0.0001 &   11.662 &      0.025 & J99,H00  & K+ \\
SN~1999cp &    51364.2 &     0.0113 & 0.0001 &   14.741 &      0.039 & K00   & K+ \\
SN~1999ee &    51470.1 &     0.0102 & 0.0001 &   14.948 &      0.017 & K04a  & K+ \\
SN~1999ek &    51482.5 &     0.0176 & 0.0001 &   15.885 &      0.027 & K04b  & K+ \\
SN~1999gp &    51550.7 &     0.0258 & 0.0001 &   16.722 &      0.093 & K01   & K+ \\
SN~2000E  &    51577.5 &     0.0045 & 0.0001 &   13.516 &      0.033 & V03   & K+ \\
SN~2000bh &    51634.5 &     0.0246 & 0.0001 &   16.541 &      0.054 & K04a  & K+ \\
SN~2000bk &    51645.7 &     0.0285 & 0.0001 &   17.151 &      0.072 & K01   & K+ \\
SN~2000ca &    51667.7 &     0.0251 & 0.0001 &   16.556 &      0.048 & K04a  & K+ \\
SN~2000ce &    51670.6 &     0.0169 & 0.0001 &   15.878 &      0.094 & K01   & K+ \\
SN~2001ba &    52035.3 &     0.0312 & 0.0001 &   17.212 &      0.034 & K04a  & K+ \\
SN~2001bt &    52064.1 &     0.0144 & 0.0001 &   15.643 &      0.030 & K04a  & K+ \\
SN~2001cn &    52072.6 &     0.0154 & 0.0001 &   15.591 &      0.053 & K04b  & K+ \\
SN~2001cz &    52104.9 &     0.0170 & 0.0001 &   15.603 &      0.053 & K04b  & K+ \\
SN~2001el &    52182.3 &     0.0036 & 0.0001 &   12.871 &      0.025 & K03   & K+ \\
SN~2002bo &    52357.3 &     0.0057 & 0.0001 &   13.822 &      0.026 & K04b  & K+ \\
SN~2002dj &    52450.8 &     0.0113 & 0.0001 &   14.669 &      0.021 & P08   & K+ \\
SN~2003du &    52768.2 &     0.0074 & 0.0001 &   14.417 &      0.050 & St07  & K+ \\
SN~2004S  &    53040.2 &     0.0100 & 0.0001 &   14.693 &      0.040 & K07   & K+ \\
SN~2004ef &    53264.5 &     0.0294 & 0.0001 &   17.208 &      0.128 & C10   & CSP \\
SN~2004eo &    53278.5 &     0.0146 & 0.0001 &   15.692 &      0.043 & Pa07b,C10  & CSP \\
SN~2004ey &    53304.9 &     0.0143 & 0.0001 &   15.672 &      0.022 & C10 & CSP \\
SN~2004gs &    53354.7 &     0.0280 & 0.0001 &   17.369 &      0.122 & C10  & CSP \\
SN~2004gu &    53366.1 &     0.0477 & 0.0001 &   17.995 &      0.071 & C10  & CSP \\
SN~2005M  &    53406.2 &     0.0236 & 0.0001 &   16.570 &      0.022 & C10  & CSP \\
SN~2005ag &    53415.1 &     0.0806 & 0.0001 &   18.980 &      0.083 & C10  & CSP \\
SN~2005al &    53430.1 &     0.0140 & 0.0001 &   15.749 &      0.064 & C10  & CSP \\
SN~2005am &    53435.1 &     0.0097 & 0.0001 &   14.144 &      0.056 & C10  & CSP \\
SN~2005ao &    53441.2 &     0.0384 & 0.0001 &   17.805 &      0.075 & WV08  & WV08 \\
SN~2005cf &    53534.0 &     0.0067 & 0.0001 &   13.914 &      0.018 & WV08,Pa07a  & WV08 \\
SN~2005ch &    53535.0 &     0.0285 & 0.0001 &   16.996 &      0.066 & WV08  & WV08 \\
SN~2005el &    53648.2 &     0.0148 & 0.0001 &   15.647 &      0.039 & WV08,C10  & WV08 \\
SN~2005eq &    53655.9 &     0.0279 & 0.0001 &   17.159 &      0.042 & WV08,C10  & WV08 \\
SN~2005eu &    53665.8 &     0.0337 & 0.0001 &   17.167 &      0.066 & WV08  & WV08 \\
SN~2005hc &    53668.2 &     0.0444 & 0.0001 &   17.929 &      0.063 & C10  & CSP \\
SN~2005hj &    53675.8 &     0.0564 & 0.0001 &   18.338 &      0.119 & S11  & CSP \\
SN~2005iq &    53687.4 &     0.0323 & 0.0001 &   17.603 &      0.054 & WV08,C10  & WV08 \\
SN~2005kc &    53698.2 &     0.0134 & 0.0001 &   15.555 &      0.024 & C10  & CSP \\
SN~2005ki &    53705.8 &     0.0211 & 0.0001 &   16.359 &      0.051 & C10  & CSP \\
SN~2005na &    53741.3 &     0.0270 & 0.0001 &   16.829 &      0.040 & WV08,C10   & WV08 \\
SN~2006D  &    53757.0 &     0.0085 & 0.0001 &   14.585 &      0.028 & WV08,C10   & WV08 \\
SN~2006N  &    53759.2 &     0.0145 & 0.0001 &   16.132 &      0.118 & WV08  & WV08 \\
SN~2006ac &    53781.2 &     0.0247 & 0.0001 &   16.725 &      0.065 & WV08  & WV08 \\
SN~2006ax &    53827.5 &     0.0187 & 0.0001 &   15.971 &      0.021 & WV08,C10 & WV08 \\
SN~2006bh &    53833.4 &     0.0104 & 0.0001 &   15.058 &      0.059 & C10  & CSP \\
SN~2006br &    53851.4 &     0.0263 & 0.0001 &   17.112 &      0.084 & S11  & CSP \\
SN~2006cp &    53897.2 &     0.0241 & 0.0001 &   16.740 &      0.108 & WV08  & WV08 \\
SN~2006ej &    53975.1 &     0.0188 & 0.0001 &   16.397 &      0.069 & S11  & CSP \\
SN~2006eq &    53971.4 &     0.0480 & 0.0001 &   18.564 &      0.292 & C10  & CSP \\
SN~2006et &    53994.7 &     0.0210 & 0.0001 &   16.288 &      0.021 & S11  & CSP \\
SN~2006ev &    53987.4 &     0.0272 & 0.0001 &   17.346 &      0.072 & S11  & CSP \\
SN~2006gj &    53998.3 &     0.0274 & 0.0001 &   17.169 &      0.190 & S11  & CSP \\
SN~2006gr &    54012.9 &     0.0331 & 0.0001 &   18.052 &      0.274 & WV08  & WV08 \\
SN~2006gt &    54000.1 &     0.0431 & 0.0001 &   18.226 &      0.254 & C10  & CSP \\
SN~2006hb &    53997.3 &     0.0152 & 0.0001 &   15.828 &      0.107 & S11  & CSP \\
SN~2006hx &    54022.6 &     0.0438 & 0.0001 &   17.817 &      0.055 & S11  & CSP \\
SN~2006is &    53996.1 &     0.0313 & 0.0001 &   17.016 &      0.219 & S11  & CSP \\
SN~2006kf &    54040.4 &     0.0205 & 0.0001 &   16.497 &      0.086 & S11  & CSP \\
SN~2006le &    54048.1 &     0.0174 & 0.0001 &   16.234 &      0.023 & WV08  & WV08 \\
SN~2006lf &    54045.7 &     0.0130 & 0.0001 &   15.265 &      0.042 & WV08  & WV08 \\
SN~2006lu &    54037.9 &     0.0548 & 0.0001 &   17.693 &      0.219 & S11  & CSP \\
SN~2006ob &    54062.0 &     0.0577 & 0.0001 &   18.761 &      0.194 & S11  & CSP \\
SN~2006os &    54064.6 &     0.0317 & 0.0001 &   17.326 &      0.052 & S11  & CSP \\
SN~2007A  &    54113.9 &     0.0160 & 0.0001 &   15.957 &      0.049 & S11  & CSP \\
SN~2007S  &    54145.4 &     0.0158 & 0.0001 &   15.489 &      0.020 & S11  & CSP \\
SN~2007af &    54174.8 &     0.0075 & 0.0001 &   13.613 &      0.013 & S11  & CSP \\
SN~2007ai &    54174.8 &     0.0324 & 0.0001 &   17.078 &      0.036 & S11  & CSP \\
SN~2007as &    54181.3 &     0.0180 & 0.0001 &   16.119 &      0.047 & S11  & CSP \\
SN~2007bc &    54201.3 &     0.0226 & 0.0001 &   16.514 &      0.056 & S11  & CSP \\
SN~2007bd &    54207.6 &     0.0322 & 0.0001 &   17.343 &      0.052 & S11  & CSP \\
SN~2007ca &    54228.5 &     0.0159 & 0.0001 &   15.666 &      0.029 & S11  & CSP \\
SN~2007cq &    54280.6 &     0.0246 & 0.0001 &   16.998 &      0.102 & WV08  & WV08 \\
SN~2007jg &    54366.6 &     0.0362 & 0.0001 &   17.873 &      0.051 & S11  & CSP \\
SN~2007le &    54399.8 &     0.0051 & 0.0001 &   13.922 &      0.013 & S11  & CSP \\
SN~2007nq &    54396.5 &     0.0433 & 0.0001 &   18.008 &      0.141 & S11  & CSP \\
SN~2007on &    54419.8 &     0.0060 & 0.0001 &   13.293 &      0.092 & S11  & CSP  \\
SN~2008C  &    54466.6 &     0.0173 & 0.0001 &   16.062 &      0.043 & S11  & CSP \\
SN~2008R &     54490.6 &     0.0125 & 0.0001 &   15.547 &      0.205 & S11  & CSP \\
SN~2008bc &    54550.7 &     0.0160 & 0.0001 &   15.744 &      0.023 & S11  & CSP \\
SN~2008bq &    54564.6 &     0.0345 & 0.0001 &   17.523 &      0.129 & S11  & CSP \\
SN~2008fp &    54731.7 &     0.0067 & 0.0001 &   13.507 &      0.014 & S11  & CSP \\
SN~2008gp &    54779.9 &     0.0324 & 0.0001 &   17.359 &      0.082 & S11  & CSP \\
SN~2008hv &    54817.6 &     0.0143 & 0.0001 &   15.541 &      0.046 & S11  & CSP \\
SN~2008ia &    54813.0 &     0.0225 & 0.0001 &   16.477 &      0.066 & S11  & CSP \\
PTF09dlc &    55073.7 &     0.0662 & 0.0001 &   18.995 &      0.046 & BN12  & BN12 \\
PTF10hdv &    55344.1 &     0.0548 & 0.0001 &   18.608 &      0.016 & BN12  & BN12 \\
PTF10hmv &    55351.4 &     0.0333 & 0.0001 &   17.534 &      0.018 & BN12  & BN12 \\
PTF10mwb &    55390.7 &     0.0315 & 0.0001 &   17.412 &      0.066 & BN12  & BN12 \\
PTF10ndc &    55390.3 &     0.0820 & 0.0001 &   19.402 &      0.036 & BN12  & BN12 \\
PTF10nlg &    55391.5 &     0.0562 & 0.0001 &   18.655 &      0.040 & BN12  & BN12 \\
PTF10qyx &    55426.1 &     0.0647 & 0.0001 &   19.125 &      0.024 & BN12  & BN12 \\
PTF10tce &    55442.0 &     0.0392 & 0.0001 &   18.045 &      0.023 & BN12  & BN12 \\
PTF10ufj &    55456.5 &     0.0758 & 0.005 &   19.307 &      0.035 & BN12  & BN12 \\
PTF10wnm &    55476.5 &     0.0640 & 0.0001 &   18.969 &      0.019 & BN12  & BN12 \\
PTF10wof &    55474.2 &     0.0508 & 0.0001 &   18.587 &      0.020 & BN12  & BN12 \\
PTF10xyt &    55490.9 &     0.0478 & 0.0001 &   18.477 &      0.099 & BN12  & BN12 \\
PTF11qmo &  55894 & 0.05696  &  0.0001 & 18.503 & 0.188  & W13  & W13 \\
PTF11qpc &  55902 & 0.09084  &  0.0001 & 19.687 & 0.082  & W13  & W13 \\
PTF11qri &  55897 & 0.057    &  0.001  & 18.689 & 0.147  & W13  & W13 \\
PTF11qzq &  55905 & 0.06     &  0.01   & 18.634 & 0.383  & W13  & W13 \\
SN~2011fs & 55833 & 0.01958  &  0.0001 & 16.141 & 0.085  & W13  & W13 \\
SN~2011gf & 55827 & 0.02626  &  0.0001 & 16.841 & 0.010  & W13  & W13 \\
SN~2011gy & 55865 & 0.01623  &  0.0001 & 15.630 & 0.194  & W13  & W13 \\
SN~2011ha & 55842 & 0.093    &  0.001  & 20.067 & 0.214  & W13  & W13 \\
SN~2011hb & 55872 & 0.02715  &  0.0001 & 17.026 & 0.068  & W13  & W13 \\
SN~2011hr & 55883 & 0.01453  &  0.0001 & 15.022 & 0.200  & W13  & W13 \\
SN~2011io & 55894 & 0.04     &  0.01   & 17.841 & 0.560  & W13  & W13 \\
SN~2011iu & 55894 & 0.04475  &  0.0001 & 18.005 & 0.169  & W13  & W13 \\
\enddata
\tablenotetext{a}{$t_{\rm max}$ from $B$-band optical light curve fits using SNooPy for WV08 and CSP and reported $B$-band $t_{\rm max}$ from \citet{Maguire12} for BN12.}  
\tablenotetext{b}{Reference codes  J99: \citet{Jha99}; H00: \citet{Hernandez00}; K00: \citet{Krisciunas00}; K04a: \citet{Krisciunas04a}; K04b: \citet{Krisciunas04b}; Ph06: \citet{Phillips06}; Pa07a: \citet{Pastorello07a}; Pa07b: \citet{Pastorello07b}; St07: \citet{Stanishev07}; WV08: \citet{Wood-Vasey08}; C10: \citet{Contreras10}; S11: \citet{Stritzinger11}; BN12: \citet{Barone-Nugent12}; W13: this present paper.}
\tablenotetext{c}{Sample name used for the divisions in the analysis.  Some SNe~Ia were observed by multiple projects.  We assign each SNe~Ia to a single sample for the purposes of quoting dispersions and distributions in the analysis.}
\label{table:HubbleFigData}
\end{deluxetable}


\end{document}